\documentclass[11pt,titlepage]{article}
\usepackage{amssymb}
\usepackage{amsfonts}
\usepackage{amsmath}

\newtheorem{theorem}{Theorem}

\newtheorem{corollary}[theorem]{Corollary}

\newtheorem{lemma}[theorem]{Lemma}

\newtheorem{remark}[theorem]{Remark}

\newenvironment{proof}[1][Proof]{\noindent\textbf{#1.} }{\ \rule{0.5em}{0.5em}}
\input{tcilatex}

\begin{document}

\title{Symplectically Covariant Schr\"{o}dinger Equation in Phase Space}
\author{Maurice A de Gosson \\
Universit\"{a}t Potsdam, Inst. f. Mathematik \\
Am Neuen Palais 10, D-14415 Potsdam\\
and\\
Universidade de S\~{a}o Paulo \\
Departamento de Matem\~{a}tica \\
Rua do Mat\~{a}o 1010, \\
\ CEP 05508-900 \ S\~{a}o Paulo\\
E-mail address: maurice.degosson@gmail.com}
\maketitle

\begin{abstract}
A classical theorem of Stone and von Neumann says that the Schr\"{o}dinger
representation is, up to unitary equivalences, the only irreducible
representation of the Heisenberg group on the Hilbert space of
square-integrable functions on configuration space. Using the Wigner-Moyal
transform we construct an irreducible representation of the Heisenberg group
on a certain Hilbert space of square-integrable functions defined on phase
space. This allows us to extend the usual Weyl calculus into a phase-space
calculus and leads us to a quantum mechanics in phase space, equivalent to
standard quantum mechanics. We also briefly discuss the extension of
metaplectic operators to phase space and the probabilistic interpretation of
the solutions of the phase space Schr\"{o}dinger equation
\end{abstract}

\tableofcontents

\section{Introduction and Motivations}

In a recent Letter \cite{physa} we have shortly discussed and justified the
Schr\"{o}dinger equation in phase space 
\begin{equation*}
i\hbar \frac{\partial }{\partial t}\Psi (x,p,t)=H(x+i\hbar \tfrac{\partial }{%
\partial p},-i\hbar \tfrac{\partial }{\partial x})\Psi (x,p,t)
\end{equation*}%
proposed by Torres-Vega and Frederick in \cite{TV1,TV2}, and obtained by
these authors using a generalized version of the Husimi transform. In this
paper we set sail to sketch a complete theory for the related equation 
\begin{equation*}
i\hbar \frac{\partial }{\partial t}\Psi (x,p,t)=H\left( \tfrac{1}{2}x+i\hbar 
\tfrac{\partial }{\partial p},\tfrac{1}{2}p-i\hbar \tfrac{\partial }{%
\partial x}\right) \Psi (x,p,t)\text{.}
\end{equation*}%
where the variables $x$ and $p$ are placed on equal footing. We will see
that in addition to the greater aesthetic\footnote{%
Admittedly, this is a subjective criterion!} appeal of the latter equation
it has the advantage of yielding a more straightforward physical
interpretations of its solutions.

This paper is reasonably self-contained; we have given rather detailed
proofs since there are many technicalities which are not always immediately
obvious.

\subsection{General Discussion}

One of the pillars of non-relativistic quantum mechanics is Schr\"{o}%
dinger's equation%
\begin{equation}
i\hbar \frac{\partial \psi }{\partial t}=-\frac{\hbar ^{2}}{2m}\nabla _{\vec{%
r}}^{2}\psi +V(\vec{r},t)\psi  \label{erwin1}
\end{equation}%
where the operator on the right-hand side is obtained from the Hamiltonian
function 
\begin{equation*}
H=\frac{1}{2m}\vec{p}^{2}+V(\vec{r},t)
\end{equation*}%
by replacing the momentum vector $\vec{p}$ by the operator $-i\hbar \nabla _{%
\vec{r}}$ and letting the position vector $\vec{r}$ stand as it is. But how
did Schr\"{o}dinger arrive at this equation? He arrived at it using what the
novelist Arthur Koestler called a \textquotedblleft
sleepwalker\textquotedblright\ argument, elaborating on Hamilton's
optical--mechanical analogy, and taking several mathematically illegitimate
steps (see Jammer \cite{Jammer} or Moore \cite{WMO} for a thorough
discussion of Schr\"{o}dinger's argument). In fact Schr\"{o}dinger's
equation can be rigorously justified for quadratic or linear potentials if
one uses the theory of the metaplectic group (see our discussion in \cite%
{ICP} , Chapters 6 and 7), but it \textit{cannot} be mathematically
justified for arbitrary Hamiltonian functions; it can only be made \textit{%
plausible} by using formal analogies: this is what is done in all texts on
quantum mechanics, and Dirac's treatise \cite{Dirac}, \textit{p}. 108--111)
is of course not an exception. The gist of Schr\"{o}dinger's argument,
recast in modern terms, is the following: a \textquotedblleft matter
wave\textquotedblright\ consists --as all waves do-- of an amplitude and a
phase. Consider now a particle with initial position vector $\vec{r}%
_{0}=(x(0),y(0),z(0))$. That particle moves under the influence of some
potential and its position vector becomes $\vec{r}(t)=(x(t),y(t),z(t))$ at
time $t$. The change of phase of the matter wave associated with the
particle is then \textit{postulated} to be the integral%
\begin{equation}
\Delta \Phi =\frac{1}{\hbar }\int_{\Gamma }\vec{p}\cdot \mathrm{d}\vec{r}-H%
\mathrm{d}t  \label{deltaphi}
\end{equation}%
calculated along the arc of trajectory $\Gamma $ joining the initial point $%
\vec{r}_{0}$ to the final point $\vec{r}(tt$ in space-time; $\vec{p}%
=(p_{x},p_{y},p_{z})$ is the momentum vector and $H=H(\vec{r},\vec{p},t)$
the Hamiltonian function. The choice (\ref{deltaphi}) for $\Delta \Phi $ is
dictated by the fact that it represents the variation in action when the
particle moves from its initial position to its final position. Now, in most
cases of interest the initial and final \textit{position} vectors uniquely
determine the initial and final \textit{momentum vectors} if $t$ is
sufficiently small, so that $\hbar \Delta \Phi $ can be identified with
Hamilton's principal function\textit{\ }$W(\vec{r}_{0},\vec{r},t)$ (see \cite%
{HGoldstein,ICP}), and the latter is a solution of Hamilton--Jacobi's
equation%
\begin{equation}
\frac{\partial W}{\partial t}+H(\vec{r},\nabla _{\vec{r}}W,t)=0\text{.}
\label{HJ}
\end{equation}%
Schr\"{o}dinger knew that the properties of the \textquotedblleft action
form\textquotedblright 
\begin{equation}
\mathcal{A}=\vec{p}\cdot \mathrm{d}\vec{r}-H\mathrm{d}t  \label{action1}
\end{equation}%
led to this equation, and this was all he needed to describe the
time-evolution of the phase. We now make an essential remark: the property
that $\Delta \Phi $ can be identified with Hamilton's principal function is
intimately related to the fact that the action form $\mathcal{A}$ is a 
\textit{relative integral invariant}. This means that if $\gamma $ and $%
\gamma ^{\prime }$ are two closed curves in the $(\vec{r},\vec{p},t)$ space
encircling the same tube of Hamiltonian trajectories, then we have%
\begin{equation*}
\oint\nolimits_{\gamma }\vec{p}\cdot \mathrm{d}\vec{r}-H\mathrm{d}%
t=\oint\nolimits_{\gamma ^{\prime }}\vec{p}\cdot \mathrm{d}\vec{r}-H\mathrm{d%
}t
\end{equation*}%
(this formula is a consequence of Stoke's theorem and generalizes to an
arbitrary number of dimensions; see for instance \cite{AP,ICP}).

\subsection{Other possible Schr\"{o}dinger equations}

We now make the following crucial observation, upon which much of this paper
relies: the action form $\mathcal{A}$ is not the only relative integral
invariant associated to the Hamiltonian $H$. In fact, for any real scalar $%
\lambda $ the differential form%
\begin{equation*}
\mathcal{A}_{\lambda }=\lambda \vec{p}\cdot \mathrm{d}\vec{r}+(\lambda -1)%
\vec{r}\cdot \mathrm{d}\vec{p}-H\mathrm{d}t
\end{equation*}%
also satisfies the equality%
\begin{equation*}
\oint\nolimits_{\gamma }\mathcal{A}_{\lambda }=\oint\nolimits_{\gamma
^{\prime }}\mathcal{A}_{\lambda }
\end{equation*}%
and is hence also a relative integral invariant. This is immediately checked
by noting that since $\gamma $ is a closed curve we have%
\begin{equation*}
\oint\nolimits_{\gamma }\vec{p}\cdot \mathrm{d}\vec{r}+\vec{r}\cdot \mathrm{d%
}\vec{p}=\oint_{\gamma }\mathrm{d}(\vec{p}\cdot \vec{r})=0
\end{equation*}%
and hence%
\begin{equation*}
\oint\nolimits_{\gamma }\lambda \vec{p}\cdot \mathrm{d}\vec{r}+(\lambda -1)%
\vec{r}\cdot \mathrm{d}\vec{p}=\oint\nolimits_{\gamma }\lambda \vec{p}\cdot 
\mathrm{d}\vec{r}+(1-\lambda )\lambda \vec{p}\cdot \mathrm{d}\vec{r}%
=\oint\nolimits_{\gamma }\vec{p}\cdot \mathrm{d}\vec{r}\text{.}
\end{equation*}%
A particularly neat choice is $\lambda =1/2$; it leads to the
\textquotedblleft symmetrized action\textquotedblright 
\begin{equation}
\mathcal{A}_{1/2}=\frac{1}{2}(\vec{p}\cdot \mathrm{d}\vec{r}-\vec{r}\cdot 
\mathrm{d}\vec{p})-H\mathrm{d}t  \label{symmac}
\end{equation}%
where the position and momentum variables now play identical roles, up to
the sign.

Let us investigate the quantum-mechanical consequences of the choice $%
\lambda =1/2$. We consider the very simple situation where the Hamiltonian
function is linear in the position and momentum variables; more specifically
we assume that%
\begin{equation*}
H_{0}=\vec{p}\cdot \vec{r}_{0}-\vec{p}_{0}\cdot \vec{r}\text{.}
\end{equation*}%
The solutions of the associated equations of motion%
\begin{equation*}
\frac{\mathrm{d}}{\mathrm{d}t}\vec{r}(t)=\vec{r}_{0}\text{ \ \textit{and} \ }%
\frac{\mathrm{d}}{\mathrm{d}t}\vec{p}(t)=\vec{p}_{0}
\end{equation*}%
are the functions%
\begin{equation*}
\vec{r}(t)=\vec{r}(0)+\vec{r}_{0}t\text{ \ \textit{and} \ }\vec{p}(t)=\vec{p}%
(0)+\vec{p}_{0}t
\end{equation*}%
hence the motion is just translation in phase space in the direction of the
vector $(\vec{r}_{0},\vec{p}_{0})$. An immediate calculation shows that the
standard change in phase (\ref{deltaphi}), expressed in terms of the final
position $\vec{r}=\vec{r}(t)$, is%
\begin{equation}
\Delta \Phi =\Phi (\vec{r};t)=\frac{1}{\hbar }(t\vec{p}_{0}\cdot \vec{r}-%
\frac{t^{2}}{2}\vec{p}_{0}\cdot \vec{r}_{0});  \label{wrt}
\end{equation}%
this function of course trivially satisfies the Hamilton--Jacobi equation
for $H_{0}$. Assuming that the initial wavefunction is $\psi _{0}=\psi _{0}(%
\vec{r})$, a straightforward calculation shows that the function 
\begin{equation}
\psi (\vec{r},t)=\exp \left[ \frac{i}{\hbar }\Phi (\vec{r};t)\right] \psi
_{0}(\vec{r}-t\vec{r}_{0})  \label{wf1}
\end{equation}%
is a solution of the standard Schr\"{o}dinger equation%
\begin{equation*}
i\hbar \frac{\partial \psi }{\partial t}=(-i\hbar \vec{r}_{0}\cdot \nabla _{%
\vec{r}}-\vec{p}_{0}\cdot \vec{r})\psi =H_{0}(\vec{r},-i\hbar \nabla
_{r})\psi \text{.}
\end{equation*}%
Suppose now that instead of using definition (\ref{deltaphi}) for the change
in phase we use instead the modified action associated with $\mathcal{A}%
_{1/2}$. Then 
\begin{equation}
\Delta \Phi _{1/2}=\frac{1}{\hbar }\int_{\Gamma }\tfrac{1}{2}(\vec{p}\cdot 
\mathrm{d}\vec{r}-\vec{r}\cdot \mathrm{d}\vec{p})-H\mathrm{d}t\text{;}
\label{delfla}
\end{equation}%
integrating and replacing $\vec{r}(0)$ with $\vec{r}-\vec{r}_{0}t$ and $\vec{%
p}(0)$ with $\vec{p}-\vec{p}_{0}t$ this leads to the expression 
\begin{equation*}
\Phi _{1/2}(\vec{r},\vec{p};t)=\frac{t}{2}(\vec{p}\cdot \vec{r}_{0}-\vec{p}%
_{0}\cdot \vec{r})
\end{equation*}%
which, in addition to time, depends on both $\vec{r}$ and $\vec{p}$; it is
thus defined on \textit{phase space}, and not on configuration space as was
the case for (\ref{wrt}). The function $\Phi _{1/2}(\vec{r},\vec{p};t)$ does
not verify the ordinary Hamilton-Jacobi equation (\ref{HJ}); it however
verifies its symmetrized variant%
\begin{equation}
\frac{\partial \Phi _{1/2}}{\partial t}+H_{0}\left( \tfrac{1}{2}\vec{r}%
+\nabla _{\vec{p}}\Phi _{1/2},\tfrac{1}{2}\vec{p}-\nabla _{\vec{r}}\Phi
_{1/2}\right) =0  \label{HJS}
\end{equation}%
as is checked by a straightforward calculation. This property opens the
gates to \textit{quantum mechanics in phase space}: assume again that we
have an initial wavefunction $\psi _{0}=\psi _{0}(\vec{r})$ and set%
\begin{equation}
\Psi (\vec{r},\vec{p},t)=\exp \left[ \frac{i}{\hbar }\Phi _{1/2}(\vec{r};t)%
\right] \psi _{0}(\vec{r}-t\vec{r}_{0})\text{.}  \label{wf2}
\end{equation}%
Using (\ref{HJS}) one finds that 
\begin{equation}
i\hbar \frac{\partial \Psi }{\partial t}=\widehat{H_{0}}\left( \tfrac{1}{2}%
\vec{r}+i\hbar \nabla _{\vec{p}},\tfrac{1}{2}\vec{p}-i\hbar \nabla _{\vec{r}%
}\right) \Psi \text{;}  \label{erwin2}
\end{equation}%
there is in fact no reason to assume that the initial wavefunction depends
only on $\vec{r}$; choosing $\Psi _{0}=\Psi _{0}(\vec{r},\vec{p})$ the same
argument shows that the function%
\begin{equation}
\Psi (\vec{r},\vec{p},t)=\exp \left[ \frac{i}{\hbar }\Phi _{1/2}(\vec{r};t)%
\right] \Psi _{0}(\vec{r}-t\vec{r}_{0},\vec{p}-t\vec{p}_{0})  \label{psprt}
\end{equation}%
is a solution of (\ref{erwin2}) with initial condition $\Psi _{0}$. Observe
that the operator $\widehat{H_{0}}$ in the \textquotedblleft phase-space Schr%
\"{o}dinger equation\textquotedblright\ (\ref{erwin2}) is obtained from the
Hamiltonian function $H_{0}$ using the phase space quantization rule%
\begin{equation*}
x\longrightarrow \widehat{X}=\tfrac{1}{2}x+i\hbar \frac{\partial }{\partial
p_{x}}\text{ \ , \ }p_{x}\longrightarrow \widehat{P}_{x}=\tfrac{1}{2}%
p_{x}-i\hbar \frac{\partial }{\partial x}
\end{equation*}%
and similar rules for the $y,z$ variables. The operators $\widehat{X},%
\widehat{P}_{x}$, etc. obey the usual canonical commutation relations:%
\begin{equation*}
\lbrack \widehat{X},\widehat{P_{x}}]=-i\hbar \text{ , \ }[\widehat{Y},%
\widehat{P_{y}}]=-i\hbar \text{ , \ }[\widehat{Z},\widehat{P_{z}}]=-i\hbar
\end{equation*}%
and this suggests that these quantization rules could be consistent with the
existence of an irreducible representation of the Heisenberg group in phase
space. \ This will be proven in Section \ref{seche}, where we will
explicitly construct this representation.

The equation (\ref{erwin2}) corresponds, as we have seen to the choice $%
\lambda =1/2$ for the integral invariant $\mathcal{A}_{\lambda }$; any other
choice is \textit{per se }equally good. For instance $\lambda =1$
corresponds to the standard Schr\"{o}dinger equation; if we took $\lambda =0$
we would obtain the phase-space Schr\"{o}dinger equation%
\begin{equation}
i\hbar \frac{\partial \Psi }{\partial t}=H_{0}(\vec{r}+i\hbar \nabla _{\vec{p%
}},-i\hbar \nabla _{r})\Psi \text{. }  \label{TF}
\end{equation}%
considered by Torres-Vega and Frederick \cite{TV1,TV2} and discussed in \cite%
{physa}.

The aesthetic appeal of the Schr\"{o}dinger equation in phase space in the
form (\ref{erwin2}) is indisputable, because it reinstates in quantum
mechanics the symmetry of classical mechanics in its Hamiltonian formulation 
\begin{equation}
\frac{\mathrm{d}\vec{r}}{\mathrm{d}t}=\nabla _{\vec{p}}H\text{ \ , \ }\frac{%
\mathrm{d}\vec{p}}{\mathrm{d}t}=-\nabla _{\vec{r}}H\text{;}  \label{hamil1}
\end{equation}%
in both (\ref{erwin1}) and (\ref{hamil1}) the variables $x$ and $p$ are
placed, up to a change of sign, on the same footing.

\subsection{Notations}

We will work with systems having $N$ degrees of freedom; we denote the
position vector of such a system by $x=(x_{1},...,x_{N})$ and its momentum
vector by $p=(p_{1},...,p_{N})$. \ We will also use the collective notation $%
z=(x,p)$ for the generic phase space variable. Configuration space is
denoted by $\mathbb{R}_{x}^{N}$ and phase space by $\mathbb{R}_{z}^{2N}$.
The generalized gradients in $x$ and $p$ are written%
\begin{equation*}
\frac{\partial }{\partial x}=\left( \frac{\partial }{\partial x_{1}},...,%
\frac{\partial }{\partial x_{N}}\right) \text{ , }\frac{\partial }{\partial p%
}=\left( \frac{\partial }{\partial p_{1}},...,\frac{\partial }{\partial p_{N}%
}\right) \text{.}
\end{equation*}%
For reasons of notational economy we will write $Mu^{2}$ instead of $Mu\cdot
u$ when $M$ is a matrix and $u$ a vector.

We denote by $z\wedge z^{\prime}$ the symplectic product of $z=(x,p)$, $%
z^{\prime}=(x^{\prime},p^{\prime})$:%
\begin{equation*}
z\wedge z^{\prime}=p\cdot x^{\prime}-p^{\prime}\cdot x
\end{equation*}
where the dot $\cdot$ is the usual (Euclidean) scalar product. In matrix
notation:%
\begin{equation*}
z\wedge z^{\prime}=(z^{\prime})^{T}Jz\text{ \ , \ }J=%
\begin{bmatrix}
0 & I \\ 
-I & 0%
\end{bmatrix}%
\end{equation*}
where $J$ is the standard symplectic matrix $(0$ and $I$ are the $N\times N$
zero and identity matrices). We denote by $Sp(N)$ the symplectic group of
the $(x,p)$ phase space: it consists of all real $2N\times2N$ matrices $S$
such that $Sz\wedge Sz^{\prime}=z\wedge z^{\prime}$; equivalently%
\begin{equation*}
S^{T}JS=SJS^{T}=J\text{.}
\end{equation*}

We denote by $(\cdot ,\cdot )$ the $L^{2}$-norm of functions on
configuration $\mathbb{R}_{x}^{N}$ and by $((\cdot ,\cdot ))$ that of
functions on phase space $\mathbb{R}_{z}^{2N}$. The corresponding norms are
denoted by $||\cdot ||$ and $|||\cdot |||$.

$\ \mathcal{S}(\mathbb{R}^{m})$ is the Schwartz space of rapidly decreasing
functions on $\mathbb{R}^{m}$ and we denote by $F$ the unitary Fourier
transform defined on $L^{2}(\mathbb{R}_{x}^{N})$ by%
\begin{equation}
F\psi (p)=\left( \tfrac{1}{2\pi \hbar }\right) ^{N/2}\int e^{-\frac{i}{\hbar 
}p\cdot x}\psi (x)\mathrm{d}^{N}x\text{.}  \label{ufo}
\end{equation}

\section{The Wigner Wave-Packet Transform\label{secwpt}}

In what follows $\phi $ will be be a rapidly decreasing function normalized
to unity: 
\begin{equation}
\phi \in \mathcal{S}(\mathbb{R}_{x}^{N})\text{ \ , \ }||\phi ||_{L^{2}(%
\mathbb{R}_{x}^{N})}^{2}=1\text{.}  \label{norm}
\end{equation}

\subsection{Definition and relation with the Wigner-Moyal transform}

We associate to $\phi $ the integral operator $U_{\phi }:L^{2}(\mathbb{R}%
_{x}^{N}\mathcal{)}\longrightarrow L^{2}(\mathbb{R}_{z}^{2N}\mathcal{)}$
defined by 
\begin{equation}
U_{\phi }\psi (z)=\left( \tfrac{1}{2\pi \hbar }\right) ^{N/2}e^{\frac{i}{%
2\hbar }p\cdot x}\int e^{-\frac{i}{\hbar }p\cdot x^{\prime }}\psi (x^{\prime
})\phi (x-x^{\prime })\mathrm{d}^{N}x^{\prime }  \label{ufi}
\end{equation}%
and we call $U_{\phi }$ the \textquotedblleft Wigner wave-packet
transform\textquotedblright\ associated with $\phi $. This terminology is
justified by the fact that the operator $U_{\phi }$ is easily expressed in
terms of the Wigner--Moyal transform 
\begin{equation}
W(\psi ,\overline{\phi })(x,p)=\left( \tfrac{1}{2\pi \hbar }\right)
^{N}\int_{\mathbb{R}^{N}}e^{-\tfrac{i}{\hbar }p\cdot y}\psi (x+\tfrac{1}{2}y)%
\overline{\phi }(x-\tfrac{1}{2}y)\mathrm{d}^{N}y  \label{wm}
\end{equation}%
of the pair $(\psi ,\overline{\phi })$ (see \cite{Folland,Littlejohn}). In
fact, performing the change of variable $x^{\prime }=\frac{1}{2}(x+y)$ in (%
\ref{ufi}) we get 
\begin{equation*}
U_{\phi }\psi (z)=\left( \frac{1}{2\pi \hbar }\right) ^{N/2}2^{-N}\int e^{-%
\frac{i}{2\hbar }p\cdot y}\psi (\tfrac{1}{2}(x+y))\phi (\tfrac{1}{2}(x-y))%
\mathrm{d}^{N}y
\end{equation*}%
that is 
\begin{equation}
U_{\phi }\psi (z)=\left( \tfrac{\pi \hbar }{2}\right) ^{N/2}W(\psi ,%
\overline{\phi })(\tfrac{1}{2}z)\text{.}  \label{defwpt}
\end{equation}

\begin{remark}
A standard --but by no means mandatory-- choice is to take for $\phi $ the
real Gaussian%
\begin{equation}
\phi _{\hbar }(x)=\left( \frac{1}{\pi \hbar }\right) ^{N/4}\exp \left( -%
\frac{1}{2\hbar }|x|^{2}\right) \text{;}  \label{fizero}
\end{equation}%
the corresponding operator $U_{\phi }$ is then (up to an exponential factor)
the \textquotedblleft coherent state representation\textquotedblright\
familiar to quantum physicists.
\end{remark}

\subsection{The fundamental property}

The interest of the Wigner wave-packet transform $U_{\phi }$ comes from the
fact that it is an isometry of $L^{2}(\mathbb{R}_{x}^{N})$ onto a closed
subspace $\mathcal{H}_{\phi }$ of $L^{2}(\mathbb{R}_{z}^{2N})$ and that it
takes the operators $x$ and $-i\hbar \partial /\partial x$ into the
operators $x/2+i\hbar \partial /\partial p$ and $p/2-i\hbar \partial
/\partial x$:

\begin{theorem}
\label{un}The Wigner wave-packet transform $U_{\phi }$ has the following
properties: (i) $U_{\phi }$ is an isometry: the Parseval formula%
\begin{equation}
((U_{\phi }\psi ,U_{\phi }\psi ^{\prime }))=(\psi ,\psi ^{\prime })
\label{parseval}
\end{equation}%
holds for all $\psi ,\psi ^{\prime }\in \mathcal{S}(\mathbb{R}_{x}^{N})$. In
particular $U_{\phi }^{\ast }U_{\phi }=I$ on\ $L^{2}(\mathbb{R}_{x}^{N})$.
(ii) The range $\mathcal{H}_{\phi }$ of $U_{\phi }$ is closed in $L^{2}(%
\mathbb{R}_{z}^{2N})$ (and is hence a Hilbert space), and the operator $%
P_{\phi }=U_{\phi }U_{\phi }^{\ast }$ is the orthogonal projection in $L^{2}(%
\mathbb{R}_{z}^{2N})$ onto $\mathcal{H}_{\phi }$. (iii) The following
intertwining relations%
\begin{equation}
\left( \frac{x}{2}+i\hbar \frac{\partial }{\partial p}\right) U_{\phi }\psi
=U_{\phi }(x\psi )\text{ \ , \ }\left( \frac{p}{2}-i\hbar \frac{\partial }{%
\partial x}\right) U_{\phi }\psi =U_{\phi }(-i\hbar \frac{\partial }{%
\partial x}\psi ).  \label{erwin4}
\end{equation}%
hold for $\psi \in \mathcal{S}(\mathbb{R}_{x}^{N})$.
\end{theorem}

\begin{proof}
\textit{(i)} Formula (\ref{parseval}) is an immediate consequence of the
property%
\begin{equation}
((W(\psi ,\phi ),W(\psi ^{\prime },\phi ^{\prime })))=\left( \tfrac{1}{2\pi
\hbar }\right) ^{N}(\psi ,\psi ^{\prime })\overline{(\phi ,\phi ^{\prime })}
\label{important}
\end{equation}%
of the Wigner--Moyal transform (see \textit{e.g. }Folland\cite{Folland} 
\textit{p}. 56; beware of the fact that Folland uses normalizations
different from ours). In fact, taking $\phi =\phi ^{\prime }$ we have%
\begin{align*}
((U_{\phi }\psi ,U_{\phi }\psi ^{\prime }))& =\left( \tfrac{\pi \hbar }{2}%
\right) ^{N}\int W(\psi ,\overline{\phi })(\tfrac{1}{2}z)\overline{W(\psi
^{\prime },\overline{\phi })}(\tfrac{1}{2}z)\mathrm{d}^{2N}z \\
& =\left( 2\pi \hbar \right) ^{N}((W(\psi ,\overline{\phi }),W(\psi ^{\prime
},\overline{\phi }))) \\
& =(\psi ,\psi ^{\prime })(\phi ,\phi )
\end{align*}%
which is formula (\ref{parseval}) since $\phi $ is normalized. To prove 
\textit{(ii)} we note that $P_{\phi }^{\ast }=P_{\phi }$ and 
\begin{equation*}
P_{\phi }^{2}=U_{\phi }(U_{\phi }^{\ast }U_{\phi })U_{\phi }^{\ast }=U_{\phi
}^{\ast }U_{\phi }=P_{\phi }
\end{equation*}%
hence $P_{\phi }$ is indeed an orthogonal projection. Let us show that its
range is $\mathcal{H}_{\phi }$; the closedness of $\mathcal{H}_{\phi }$ will
follow since the range of a projection in a Hilbert space always is closed.
Since $U_{\phi }^{\ast }U_{\phi }=I$ on $L^{2}(\mathbb{R}_{x}^{N})$ we have $%
U_{\phi }^{\ast }U_{\phi }\psi =\psi $ for every $\psi $ in $L^{2}(\mathbb{R}%
_{x}^{N})$ and hence the range of $U_{\phi }^{\ast }$ is $L^{2}(\mathbb{R}%
_{x}^{N})$. It follows that the range of $U_{\phi }$ is that of $U_{\phi
}U_{\phi }^{\ast }=P_{\phi }$ and we are done. \textit{(iii)} The
verification of the formulae (\ref{erwin4}) is purely computational, using
differentiations and partial integrations; it is therefore left to the
reader.
\end{proof}

The intertwining formulae (\ref{erwin4}) show that the Wigner wave-packet
transform takes the usual quantization rules $x\longrightarrow x$, $%
x\longrightarrow -i\hbar \frac{\partial }{\partial x}$ leading to the
standard Schr\"{o}dinger equation to the phase-space quantization rules%
\begin{equation*}
x\longrightarrow \frac{1}{2}x+i\hbar \frac{\partial }{\partial p}\text{ \ ,
\ }p\longrightarrow \frac{1}{2}p-i\hbar \frac{\partial }{\partial x}\text{;}
\end{equation*}%
observe that these rules are independent of the choice of $\phi $ and that
these rules are thus a common features of all the transforms $U_{\phi }$.

\subsection{The range of $U_{\protect\phi }$}

One should be aware of the fact that the Hilbert space $\mathcal{H}_{\phi }$
is smaller than $L^{2}(\mathbb{R}_{z}^{2N})$. This is intuitively clear
since functions in $L^{2}(\mathbb{R}_{z}^{2N})$ depend on twice as many
variables as those in $L^{2}(\mathbb{R}_{x}^{N})$ of which $\mathcal{H}%
_{\phi }$ is an isometric copy. Let us discuss this in some detail.

\begin{theorem}
\label{theorange}(i) The range of the Wigner wave-packet transform $U_{\phi
_{\hbar }}$ associated to the Gaussian (\ref{fizero}) consists of the
functions $\Psi \in L^{2}(\mathbb{R}_{z}^{2N})$ for which the conditions%
\begin{equation}
\left( \frac{\partial }{\partial x_{j}}-i\frac{\partial }{\partial p_{j}}%
\right) \left[ \exp \left( \frac{1}{2\hbar }|z|^{2}\right) \Psi (z)\right] =0%
\text{ \ for \ }1\leq j\leq N  \label{CR}
\end{equation}%
hold. (ii) For every $\phi $ the range of the Wigner wave-packet transform $%
U_{\phi }$ is isometric to $\mathcal{H}_{\phi _{\hbar }}$.
\end{theorem}

\begin{proof}
\textit{(i)} We have $U_{\phi _{\hbar }}=e^{-\frac{i}{2\hbar }p\cdot
x}V_{\phi _{\hbar }}$ where the operator $V_{\phi _{\hbar }}$ is defined by%
\begin{equation*}
V_{\phi }\psi (z)=\left( \tfrac{1}{2\pi \hbar }\right) ^{N/2}\int e^{-\tfrac{%
i}{\hbar }p\cdot x^{\prime }}\phi (x-x^{\prime })\psi (x^{\prime })\mathrm{d}%
^{N}x^{\prime }
\end{equation*}%
It is shown in\cite{Naza} that the range of $V_{\phi _{\hbar }}$ consists of
all $\Psi \in L^{2}(\mathbb{R}_{z}^{2N})$ such that%
\begin{equation*}
\left( \frac{\partial }{\partial x_{j}}-i\frac{\partial }{\partial p_{j}}%
\right) \left[ \exp \left( \frac{1}{2\hbar }|p|^{2}\right) \Psi (z)\right] =0%
\text{ \ for \ }1\leq j\leq N\text{.}
\end{equation*}%
That the range of $U_{\phi _{\hbar }}$ is characterized by (\ref{CR})
follows by an immediate calculation that is left to the reader. \textit{(ii)}
If $U_{\phi _{1}}$ and $U_{\phi _{2}}$ are two Wigner wave-packet transforms
corresponding to the choices $\phi _{1}$, $\phi _{2}$ then $U_{\phi
_{2}}U_{\phi _{1}}^{\ast }$ is an isometry of $\mathcal{H}_{\phi _{1}}$ onto 
$\mathcal{H}_{\phi _{2}}$.
\end{proof}

The result above leads us to address the following more precise question:
given $\Psi \in L^{2}(\mathbb{R}_{z}^{2N})$, can we find $\phi $ and $\psi $
in $L^{2}(\mathbb{R}_{x}^{N})$ such that $\Psi =U_{\phi }\psi $? We are
going to see that the answer is \textit{no}. Intuitively speaking the reason
is the following: if $\Psi $ is too \textquotedblleft
concentrated\textquotedblright\ in phase space, it cannot correspond via the
inverse transform $U_{\phi }^{-1}=U_{\phi }^{\ast }$ to a solution of the
standard Schr\"{o}dinger equation, because the uncertainty principle would
be violated. Let us make this precise when the function $\Psi $ is a
Gaussian. We first make the following obvious remark: in view of condition (%
\ref{CR}) every Gaussian%
\begin{equation*}
\Psi _{0}(z)=\lambda \exp \left( \frac{1}{2\hbar }|z|^{2}\right) \text{ \ ,
\ \ }\lambda \in \mathbb{C}
\end{equation*}%
is in the range of $U_{\phi _{\hbar }}$. It turns out that not only does
this particular Gaussian belong to the range of every Wigner wave-packet
transform $U_{\phi }$, but so does also the compose $\Psi _{0}\circ S$ for
every $S\in Sp(N)$:

\begin{theorem}
\label{theogauss}Let $G$ be a real positive-definite $2N\times 2N$ matrix: $%
G=G^{T}>0$. Let $\Psi _{G}\in L^{2}(\mathbb{R}_{z}^{2N})$ be the Gaussian:%
\begin{equation}
\Psi _{G}(z)=\exp \left( -\frac{1}{2\hbar }Gz^{2}\right) \text{.}
\label{psim}
\end{equation}%
(i) There exist $\psi ,\phi \in \mathcal{S}(\mathbb{R}_{x}^{N})$ such that $%
U_{\phi }\psi =\Psi _{G}$ if and only if $G\in Sp(N)$, in which case we have%
\begin{equation*}
\phi =\alpha \phi _{\hbar }\text{ \ , \ }\psi =2^{N/2}\overline{\alpha }%
\left( \pi \hbar \right) ^{N/4}\phi _{\hbar }
\end{equation*}%
where $\phi _{\hbar }$ is the Gaussian (\ref{fizero}) and $\alpha $ an
arbitrary complex constant with modulus one. (ii) Equivalently, $|\Psi
_{G}|^{2}$ must be the Wigner transform $W\psi $ of a Gaussian state 
\begin{equation}
\psi (x)=c\left( \det X\right) ^{1/4}(\pi \hbar )^{3N/4}\exp \left[ -\frac{1%
}{2\hbar }(X+iY)x^{2}\right] \text{ }  \label{argauss}
\end{equation}%
with $|c|=1$, $X$ and $Y$ real and symmetric, and $X>0$.
\end{theorem}

\begin{proof}
In view of the relation (\ref{defwpt}) between $U_{\phi }$ and the
Wigner-Moyal transform (\ref{defwpt}) is equivalent to%
\begin{equation*}
W(\psi ,\overline{\phi })(z)=\left( \frac{2}{\pi \hbar }\right) ^{N/2}\exp
\left( -\frac{2}{\hbar }Gz^{2}\right) \text{.}
\end{equation*}%
In view of Williamson's symplectic diagonalization theorem \cite{Will} there
exists $S\in Sp(N)$ such that $G=S^{T}DS$ where $D$ is the diagonal matrix%
\begin{equation*}
D=%
\begin{bmatrix}
\Lambda & 0 \\ 
0 & \Lambda%
\end{bmatrix}%
\text{ \ \ , \ \ }\Lambda =\limfunc{diag}[\lambda _{1},...,\lambda _{N}]
\end{equation*}%
the numbers $\pm i\lambda _{1},...,\lambda _{N}$, $\lambda _{j}>0$, being
the eigenvalues of $JG^{-1}$; it follows that 
\begin{equation*}
W(\psi ,\overline{\phi })(S^{-1}z)=\left( \frac{2}{\pi \hbar }\right)
^{N/2}\exp \left( -\frac{2}{\hbar }Dz^{2}\right) \text{.\ }
\end{equation*}%
In view of the metaplectic covariance property of the Wigner--Moyal
transform (see (\ref{metac1}) in Section \ref{secmetacov}) there exists a
unitary operator $\widehat{S}:\mathcal{S}(\mathbb{R}_{x}^{N})\longrightarrow 
\mathcal{S}(\mathbb{R}_{x}^{N})$ such that%
\begin{equation*}
W(\psi ,\overline{\phi })(S^{-1}z)=W(\widehat{S}\psi ,\widehat{S}\overline{%
\phi })(z)
\end{equation*}%
hence it is no restriction to assume $S=I$ and%
\begin{equation*}
W(\psi ,\overline{\phi })(z)=\left( \frac{2}{\pi \hbar }\right) ^{N/2}\exp
\left( -\frac{2}{\hbar }Dz^{2}\right) \text{.\ }
\end{equation*}%
By definition (\ref{wm}) of the Wigner-Moyal transform this is the same
thing as%
\begin{equation*}
\left( \tfrac{1}{2\pi \hbar }\right) ^{N/2}\int e^{-\frac{i}{\hbar }p\cdot
y}\psi (x+\tfrac{1}{2}y)\phi (x-\tfrac{1}{2}y)\mathrm{d}^{N}y=2^{N}\exp
\left( -\frac{2}{\hbar }Dz^{2}\right)
\end{equation*}%
that is, in view of the Fourier inversion formula,%
\begin{eqnarray*}
\psi (x+\tfrac{1}{2}y)\phi (x-\tfrac{1}{2}y) &=&2^{N}\left( \tfrac{1}{2\pi
\hbar }\right) ^{N/2}\int e^{-\frac{i}{\hbar }p\cdot y}e^{-\frac{2}{\hbar }%
Dz^{2}}\mathrm{d}^{N}p \\
&=&\left( \tfrac{2}{\pi \hbar }\right) ^{N/2}e^{-\frac{1}{\hbar }\Lambda
x^{2}}\int e^{\frac{i}{\hbar }p\cdot y}e^{-\frac{1}{\hbar }\Lambda p^{2}}%
\mathrm{d}^{N}p.
\end{eqnarray*}%
Setting $Q=2\Lambda $ in the generalized Fresnel formula%
\begin{equation*}
\left( \tfrac{1}{2\pi \hbar }\right) ^{N/2}\int e^{-\frac{i}{\hbar }p\cdot
y}e^{-\frac{1}{2\hbar }Qp^{2}}\mathrm{d}^{N}p=(\det Q)^{-1/2}e^{-\frac{1}{%
2\hbar }Q^{-1}y^{2}}
\end{equation*}%
valid for all $Q>0$ we thus have 
\begin{equation*}
\psi (x+\tfrac{1}{2}y)\phi (x-\tfrac{1}{2}y)=2^{N/2}(\det \Lambda
)^{-1/2}\exp \left[ -\frac{1}{\hbar }\left( \Lambda x^{2}+\frac{1}{4}\Lambda
y^{2}\right) \right] .
\end{equation*}%
Setting $u=x+\tfrac{1}{2}y$ and $v=x-\tfrac{1}{2}y$ this is%
\begin{multline*}
\psi (u)\phi (v)=2^{N/2}(\det \Lambda )^{-1/2}\times \\
\exp \left[ -\frac{1}{4\hbar }\left( (\Lambda +\Lambda
^{-1})(u^{2}+v^{2})+2(\Lambda -\Lambda ^{-1})u\cdot v\right) \right]
\end{multline*}%
and this is only possible if there are no terms $u\cdot v$. This condition
requires that $\Lambda =\Lambda ^{-1}$; since $\Lambda $ is
positive-definite we must have $\Lambda =I$ and hence $\Delta =I$. It
follows that%
\begin{equation*}
\psi (u)\phi (v)=2^{N/2}\exp \left[ -\frac{1}{2\hbar }(u^{2}+v^{2})\right]
\end{equation*}%
so that%
\begin{equation*}
\psi (x)\phi (0)=\psi (0)\phi (x)=2^{N/2}\exp \left( -\frac{1}{2\hbar }%
x^{2}\right) \text{.}
\end{equation*}%
It follows that both $\psi $ and $\phi $ are Gaussians of the type%
\begin{equation*}
\psi (x)=\psi (0)\exp \left( -\frac{1}{2\hbar }x^{2}\right) \text{ \ , \ }%
\phi (x)=\phi (0)\exp \left( -\frac{1}{2\hbar }x^{2}\right) ;
\end{equation*}%
since $\phi $ is normalized this requires that $\phi =\alpha \phi _{\hbar }$
with $|\alpha |=1$ and hence%
\begin{equation*}
\phi (0)=\alpha \left( \frac{1}{\pi \hbar }\right) ^{N/4}.
\end{equation*}%
Since we have $\psi (0)\phi (0)=2^{N/2}$ we must have%
\begin{equation*}
\psi (0)=\overline{\alpha }2^{N/2}\left( \pi \hbar \right) ^{N/4}
\end{equation*}%
which concludes the proof of part \textit{(i)} of the theorem. To prove 
\textit{(ii)} recall from Littlejohn \cite{Littlejohn} that the Wigner
transform of the Gaussian (\ref{argauss}) is given by the formula 
\begin{equation*}
W\psi (z)=\exp \left( -\frac{1}{\hbar }Gz^{2}\right)
\end{equation*}%
where 
\begin{equation*}
G=%
\begin{bmatrix}
X+YX^{-1}Y & YX^{-1} \\ 
X^{-1}Y & X^{-1}%
\end{bmatrix}%
\text{.}
\end{equation*}%
It is immediate to verify that $G\in Sp(N)$ and that $G$ is symmetric
positive definite. One proves \cite{CQ} that, conversely, every such $G$ can
be put in the form above, and which ends the proof of \textit{(ii)} since
the datum of $W\psi $ determines $\psi $ up to a complex factor with modulus
one.
\end{proof}

\section{Phase-Space Weyl Calculus\label{seche}}

Let $\mathbf{H}_{N}$ be the $(2N+1)$-dimensional Heisenberg group; it is
(see \textit{e.g.} \cite{Folland,Schempp}) the set of all vectors 
\begin{equation*}
(z,t)=(x_{1},...,x_{N};p_{1},...,p_{N};t)
\end{equation*}%
equipped with the multiplicative law%
\begin{equation*}
(z,t)(z^{\prime },t^{\prime })=(z+z^{\prime },t+t^{\prime }+\tfrac{1}{2}%
z\wedge z^{\prime }).
\end{equation*}

\subsection{The Schr\"{o}dinger representation}

The Schr\"{o}dinger representation of $\mathbf{H}_{N}$ is, by definition,
the mapping $\widehat{T}_{\text{Sch}}$ which to every $(z_{0},t_{0})$ in $%
\mathbf{H}_{N}$ associates the unitary operator $\widehat{T}_{\text{Sch}%
}(z_{0},t_{0})$ on $L^{2}(\mathbb{R}_{x}^{N})$ defined by 
\begin{equation}
\widehat{T}_{\text{Sch}}(z_{0},t_{0})\psi (x)=\exp \left[ \frac{i}{\hbar }%
(-t_{0}+p_{0}\cdot x-\frac{1}{2}p_{0}\cdot x_{0})\right] \psi (x-x_{0}).
\label{HG1}
\end{equation}%
A classical theorem due to Stone and von Neumann (see for instance \cite%
{Folland,Schempp} for a proof) says that the Schr\"{o}dinger representation
is irreducible (that is, no closed subspace of $L^{2}(\mathbb{R}_{x}^{N})$
other than $\{0\}$ and $L^{2}(\mathbb{R}_{x}^{N})$ are invariant by $%
\widehat{T}_{\text{Sch}}$), and that every irreducible unitary
representation of $\mathbf{H}_{N}$ is unitarily equivalent to $\widehat{T}_{%
\text{Sch}}$: if $\widehat{T}$ is another irreducible representation of $%
\mathbf{H}_{N}$ on some Hilbert space $\mathcal{H}$ then there exists a
unitary operator $U$ from $L^{2}(\mathbb{R}_{x}^{N})$ to $\mathcal{H}$ which
is bijective, and such that the following intertwining formula holds: 
\begin{equation*}
\lbrack U\circ \widehat{T}_{\text{Sch}}](z,t)=[\widehat{T}\circ U](z,t)\text{
\ for all }(z,t)\text{ in }\mathbf{H}_{N}\text{.}
\end{equation*}%
Conversely, if $U$ is a unitary operator for which this formula holds, then $%
\widehat{T}$ must be irreducible. We emphasize --heavily!-- that in the
statement above it is nowhere assumed that $\mathcal{H}$ must be $L^{2}(%
\mathbb{R}_{x}^{N})$; it can \textit{a priori} be any Hilbert space, and in
particular it can (and will be!) any of the spaces $\mathcal{H}_{\phi }$
defined in Theorem \ref{un}. We will come back to this point in a while, but
let us first recall how one passes from the Heisenberg group to the Weyl
pseudo-differential calculus. In Weyl calculus one associates to a
\textquotedblleft symbol\textquotedblright\ $A=A(x,p)$ an operator $\widehat{%
A}$ on $\mathcal{S}(\mathbb{R}_{x}^{N})$ by the formula%
\begin{equation}
\widehat{A}\psi (x)=\left( \tfrac{1}{2\pi \hbar }\right) ^{N}\iint e^{\frac{i%
}{\hbar }p\cdot (x-y)}A(\tfrac{1}{2}(x+y),p)\psi (y)\mathrm{d}^{N}y\mathrm{d}%
^{N}p\text{.}  \label{psido2}
\end{equation}%
This formula makes perfectly sense if for instance $A\in \mathcal{S}(\mathbb{%
R}_{z}^{2N})$, and one easily verifies that the \textquotedblleft Weyl
correspondence\textquotedblright\ $A\overset{\text{Weyl}}{%
\longleftrightarrow }\widehat{A}$ leads to the standard quantization rules 
\begin{equation}
x\overset{\text{Weyl}}{\longleftrightarrow }\widehat{X}=x\ \ \ \text{, }\ \ p%
\overset{\text{Weyl}}{\longleftrightarrow }\widehat{P}=-i\hbar \frac{%
\partial }{\partial x}\text{.}  \label{QR}
\end{equation}

For more general symbols the double integral must be interpreted in some
particular way. For instance, if $A$ belongs to the standard symbol class $%
S_{\rho ,\delta }^{m}(\mathbb{R}_{z}^{2N})$ with $0\leq \delta <\rho \leq 1$
that is, if for every compact $K\subset \mathbb{R}_{x}^{N}$ and all
multi-indices $\alpha ,\beta \in \mathbb{N}^{N}$ there exists $C_{K,\alpha
,\beta }$ such that%
\begin{equation*}
|\partial _{p}^{\alpha }\partial _{x}^{\beta }A(x,p)|\leq C_{K,\alpha ,\beta
}(1+|p|)^{m-\rho |\alpha |+\delta |\beta |}
\end{equation*}%
then (\ref{psido2}) should be viewed as an \textquotedblleft oscillatory
integral\textquotedblright . There are however other possible ways to
interpret this formula and make it rigorous; we refer to \cite%
{Dubin,Folland,Wong} for detailed discussions. (In particular it is shown in 
\cite{Wong} that if $\widehat{A}$ is a Hilbert--Schmidt operator if and only
if $A\in L^{2}(\mathbb{R}_{x}^{N})$).

\subsection{Heisenberg--Weyl operators}

There is another very useful way of writing Weyl operators, and this will
lead us to Weyl calculus in phase space. Setting $t_{0}=0$ in formula (\ref%
{HG1}) one obtains the so-called Heisenberg--Weyl operators $\widehat{T}_{%
\text{Sch}}(z_{0})$:%
\begin{equation}
\widehat{T}_{\text{Sch}}(z_{0})\psi (x)=\exp \left[ \frac{i}{\hbar }%
(p_{0}\cdot x-\frac{1}{2}p_{0}\cdot x_{0})\right] \psi (x-x_{0})\text{.}
\label{HW1}
\end{equation}%
It is easy to show, using Fourier analysis, that we can write the operator (%
\ref{psido2}) in the form 
\begin{equation}
\widehat{A}\psi (x)=\left( \tfrac{1}{2\pi \hbar }\right) ^{N}\int (\mathcal{F%
}_{\sigma }A)(z_{0})\widehat{T}_{\text{Sch}}(z_{0})\psi (x)\mathrm{d}%
^{2N}z_{0}  \label{weyl1}
\end{equation}%
provided that $\mathcal{F}_{\sigma }A$, the \textquotedblleft symplectic\
Fourier transform\textquotedblright\ of $A$, exists. The latter is defined
in analogy with the ordinary Fourier transform on $\mathbb{R}_{z}^{2N}$ by%
\begin{equation}
\mathcal{F}_{\sigma }A(z)=\left( \tfrac{1}{2\pi \hbar }\right) ^{N}\dint e^{-%
\frac{i}{\hbar }z\wedge z^{\prime }}A(z^{\prime })\mathrm{d}^{2N}z^{\prime }%
\text{.}  \label{asigma}
\end{equation}%
The conditions of existence of $\mathcal{F}_{\sigma }A$ are actually the
same as for the usual Fourier transform on $L^{2}(\mathbb{R}_{z}^{2N})$ to
which it reduces replacing $z$ by $-Jz$. Notice that $\mathcal{F}_{\sigma }$
is an involution: $\mathcal{F}_{\sigma }^{2}=I$.

\begin{remark}
\label{rembo}It is often convenient to write formula (\ref{weyl1}) more
economically as%
\begin{equation}
\widehat{A}=\left( \tfrac{1}{2\pi \hbar }\right) ^{N}\int (\mathcal{F}%
_{\sigma }A)(z)\widehat{T}_{\text{Sch}}(z)\mathrm{d}^{2N}z  \label{weylbo}
\end{equation}%
where the right-hand-side is interpreted as a \textquotedblleft Bochner
integral\textquotedblright , i.e. an integral with value in a Banach space.
\end{remark}

\subsection{Extension to phase space}

We now observe that when a Weyl operator is written in the form (\ref{weyl1}%
) or (\ref{weylbo}) it literally begs to be extended to phase space! In
fact, we can make the Heisenberg--Weyl operators act on functions $\Psi $ in 
$L^{2}(\mathbb{R}_{z}^{2N})$ by replacing definition (\ref{HW1}) by its
obvious extension%
\begin{equation*}
\widehat{T}_{\text{Sch}}(z_{0})\Psi (z)=\exp \left[ \frac{i}{\hbar }%
(p_{0}\cdot x-\frac{1}{2}p_{0}\cdot x_{0})\right] \Psi (z-z_{0})
\end{equation*}%
and thereafter define the action of $\widehat{A}$ on $\Psi \in \mathcal{S}(%
\mathbb{R}_{z}^{2N})$ by the formula 
\begin{equation*}
\widehat{A}\Psi (z)=\left( \tfrac{1}{2\pi \hbar }\right) ^{N}\int (\mathcal{F%
}_{\sigma }A)(z_{0})\widehat{T}_{\text{Sch}}(z_{0})\Psi (z)\mathrm{d}%
^{2N}z_{0}\text{.}
\end{equation*}

This (perfectly honest) choice would lead, using the method we will outline
in Section \ref{sesch}, to the Torres--Vega equation%
\begin{equation*}
i\hbar \frac{\partial \Psi }{\partial t}=\widehat{H}\left( x+i\hbar \frac{%
\partial }{\partial p},-i\hbar \frac{\partial }{\partial x}\right) \Psi 
\text{ }
\end{equation*}%
(see \cite{TV1,TV2}) which we discussed in \cite{physa}. Since we prefer,
for reasons of symplectic covariance, a more symmetric phase-space Schr\"{o}%
dinger equation in which $x$ and $p$ are placed on equal footing,\ we will
replace $\widehat{T}_{\text{Sch}}(z_{0})$ with the operator $\widehat{T}_{%
\text{ph}}(z_{0})$ given by%
\begin{equation}
\widehat{T}_{\text{ph}}(z_{0})\Psi (z)=\exp \left( -\frac{i}{2\hbar }z\wedge
z_{0}\right) \Psi (z-z_{0})  \label{hwnew}
\end{equation}%
(the subscript \textquotedblleft \textit{ph}\textquotedblright\ standing for
phase space) and then define the phase-space Weyl operator associated to $A$
by the formula%
\begin{equation}
\widehat{A}_{\text{ph}}\Psi (z)=\left( \tfrac{1}{2\pi \hbar }\right)
^{N}\int (\mathcal{F}_{\sigma }A)(z_{0})\widehat{T}_{\text{ph}}(z_{0})\Psi
(z)\mathrm{d}^{2N}z_{0}\text{.}  \label{weyl2}
\end{equation}%
This operator $\widehat{A}_{\text{ph}}$ acts continuously on $\mathcal{S}(%
\mathbb{R}_{z}^{2N})$ provided that $A$ is a \textit{bona fide} symbol and
can hence be extended to $L^{2}(\mathbb{R}_{z}^{2N})$ by continuity. In
accordance with the convention in Remark \ref{rembo} we will often write for
short%
\begin{equation}
\widehat{A}_{\text{ph}}=\left( \tfrac{1}{2\pi \hbar }\right) ^{N}\int (%
\mathcal{F}_{\sigma }A)(z)\widehat{T}_{\text{ph}}(z)\mathrm{d}^{2N}z
\label{bo}
\end{equation}%
where the right-hand side is again viewed as a Bochner integral.

Observe that the operators $\widehat{T}_{\text{ph}}$ satisfy the same
commutation relation as the usual Weyl--Heisenberg operators:%
\begin{equation}
\widehat{T}_{\text{ph}}(z_{1})\widehat{T}_{\text{ph}}(z_{0})=\exp \left( -%
\frac{i}{\hbar }z_{0}\wedge z_{1}\right) \widehat{T}_{\text{ph}}(z_{0})%
\widehat{T}_{\text{ph}}(z_{1})  \label{formuco2}
\end{equation}%
and we have%
\begin{equation}
\widehat{T}_{\text{ph}}(z_{0})\widehat{T}_{\text{ph}}(z_{1})=\exp \left( 
\frac{i}{2\hbar }z_{0}\wedge z_{1}\right) \widehat{T}_{\text{ph}%
}(z_{0}+z_{1}).  \label{formuco3}
\end{equation}

These properties suggest that we define the phase-space representation $%
\widehat{T}_{\text{ph}}$ of $\mathbf{H}_{N}$ in analogy with (\ref{HG1}) by
setting for $\Psi \in L^{2}(\mathbb{R}_{z}^{2N})$ 
\begin{equation}
\widehat{T}_{\text{ph}}(z_{0},t_{0})\Psi (z)=e^{\frac{i}{\hbar }t_{0}}%
\widehat{T}_{\text{ph}}(tz_{0})\Psi (z)\text{.}  \label{tps}
\end{equation}%
Clearly $\widehat{T}_{\text{ph}}(z_{0},t_{0})$ is a unitary operator in $%
L^{2}(\mathbb{R}_{z}^{2N})$; moreover a straightforward calculation shows
that%
\begin{equation}
\widehat{T}_{\text{ph}}(z_{0},t_{0})\widehat{T}_{\text{ph}}(z_{1},t_{1})=%
\widehat{T}_{\text{ph}}(z_{0}+z_{1},t_{0}+t_{1}+\tfrac{1}{2}z_{0}\wedge
z_{1})  \label{formuco1}
\end{equation}%
hence $\widehat{T}_{\text{ph}}$ is indeed a representation of the Heisenberg
group in $L^{2}(\mathbb{R}_{z}^{2N})$. We claim that the following diagram
is commutative for every Wigner wave-packet transform $U_{\phi }$:%
\begin{equation*}
\begin{array}{ccc}
L^{2}(\mathbb{R}_{x}^{N}) & \overset{U_{\phi }}{\longrightarrow } & L^{2}(%
\mathbb{R}_{z}^{2N}) \\ 
\widehat{T}_{\text{Sch}}\downarrow &  & \downarrow \widehat{T}_{\text{ph}}
\\ 
L^{2}(\mathbb{R}_{x}^{N}) & \overset{U_{\phi }}{\longrightarrow } & L^{2}(%
\mathbb{R}_{z}^{2N}).%
\end{array}%
\end{equation*}%
More precisely:

\begin{theorem}
\label{thinter}Let $U_{\phi }$ be an arbitrary Wigner wave-packet transform.
(i)We have%
\begin{equation}
\widehat{T}_{\text{ph}}(z_{0},t_{0})U_{\phi }=U_{\phi }\widehat{T}_{\text{Sch%
}}(z_{0},t_{0})  \label{unit}
\end{equation}%
hence the representation $\widehat{T}_{\text{ph}}$ is unitarily equivalent
to the Schr\"{o}dinger representation and thus an irreducible representation
of $\mathbf{H}_{N}$ on each of the Hilbert spaces $\mathcal{H}_{\phi }$.
(ii) The following intertwining formula holds for every operator $\widehat{A}%
_{\text{ph}}$: 
\begin{equation}
\widehat{A}_{\text{ph}}U_{\phi }=U_{\phi }\widehat{A}_{\text{Sch}}.
\label{inter}
\end{equation}
\end{theorem}

\begin{proof}
Proof of \textit{(i)}. It suffices to prove formula (\ref{unit}) for $%
t_{0}=0 $, that is%
\begin{equation}
\widehat{T}_{\text{ph}}(z_{0})U_{\phi }=U_{\phi }\widehat{T}_{\text{Sch}%
}(z_{0})\text{.}  \label{intertwine}
\end{equation}%
Let us write the operator $U_{\phi }$ in the form $U_{\phi }=e^{\frac{i}{%
2\hbar }p\cdot x}W_{\phi }$ where the operator $W_{\phi }$ is thus defined by%
\begin{equation}
W_{\phi }\psi (z)=\left( \tfrac{1}{2\pi \hbar }\right) ^{N/2}\int e^{-\tfrac{%
i}{\hbar }p\cdot x^{\prime }}\phi (x-x^{\prime })\psi (x^{\prime })\mathrm{d}%
^{N}x^{\prime }\text{.}  \label{vefi}
\end{equation}%
We have, by definition of $\widehat{T}_{\text{ph}}(z_{0})$ 
\begin{align*}
\widehat{T}_{\text{ph}}(z_{0})U_{\phi }\psi (z)& =\exp \left[ -\frac{i}{%
2\hbar }(z\wedge z_{0}+(p-p_{0})\cdot (x-x_{0})\right] W_{\phi }\psi
(z-z_{0}) \\
& =\exp \left[ \frac{i}{2\hbar }(-2p\cdot x_{0}+p_{0}\cdot x_{0}+p\cdot x)%
\right] W_{\phi }\psi (z-z_{0})
\end{align*}%
and, by definition of $W_{\phi }\psi $, 
\begin{align*}
W_{\phi }\psi (z-z_{0})& =\left( \tfrac{1}{2\pi \hbar }\right) ^{N/2}\int
e^{-\tfrac{i}{\hbar }(p-p_{0})\cdot x^{\prime }}\overline{\phi }(x-x^{\prime
}-x_{0})\psi (x^{\prime })\mathrm{d}^{N}x^{\prime } \\
& =\left( \tfrac{1}{2\pi \hbar }\right) ^{N/2}e^{\tfrac{i}{\hbar }%
(p-p_{0})\cdot x_{0}}\int e^{-\tfrac{i}{\hbar }(p-p_{0})\cdot x^{\prime
\prime }}\overline{\phi }(x-x^{\prime \prime })\psi (x^{\prime \prime })%
\mathrm{d}^{N}x^{\prime \prime }
\end{align*}%
where we have set $x^{\prime \prime }=x^{\prime }+x_{0}$. The overall
exponential in $\widehat{T}_{\text{ph}}(z_{0})U_{\phi }\psi (z)$ is thus%
\begin{equation*}
u_{1}=\exp \left[ \frac{i}{2\hbar }(-p_{0}\cdot x_{0}+p\cdot x-2p\cdot
x^{\prime \prime }+2p_{0}\cdot x^{\prime \prime })\right] \text{.}
\end{equation*}%
Similarly,%
\begin{multline*}
U_{\phi }(\widehat{T}_{\text{Sch}}(z_{0})\psi )(z)=\left( \tfrac{1}{2\pi
\hbar }\right) ^{N/2}e^{\frac{i}{2\hbar }p\cdot x}\times \\
\int e^{-\tfrac{i}{\hbar }p\cdot x^{\prime \prime }}\overline{\phi }%
(x-x^{\prime \prime })e^{\frac{i}{\hbar }(p_{0}\cdot x^{\prime \prime }-%
\frac{1}{2}p_{0}\cdot x_{0})}\psi (x^{\prime \prime }-x_{0})\mathrm{d}%
^{N}x^{\prime \prime }
\end{multline*}%
yielding the overall exponential%
\begin{equation*}
u_{2}=\exp \left[ \frac{i}{\hbar }\left( \frac{1}{2}p\cdot x-p\cdot
x^{\prime \prime }+p_{0}\cdot x^{\prime \prime }-\frac{1}{2}p_{0}\cdot
x_{0}\right) \right] =u_{1}
\end{equation*}%
which proves (\ref{intertwine}). The irreducibility statement follows from
Stone--von Neumann's theorem. Let us prove formula (\ref{inter}). In view of
formula (\ref{intertwine}) we have%
\begin{align*}
\widehat{A}_{\text{ph}}U_{\phi }\psi & =\left( \tfrac{1}{2\pi \hbar }\right)
^{N}\int (\mathcal{F}_{\sigma }A)(z_{0})\widehat{T}_{\text{ph}%
}(z_{0})(U_{\phi }\psi )(z)\mathrm{d}^{2N}z_{0} \\
& =\left( \tfrac{1}{2\pi \hbar }\right) ^{N}\int (\mathcal{F}_{\sigma
}A)(z_{0})U_{\phi }(\widehat{T}_{\text{Sch}}(z_{0})\psi )(z)\mathrm{d}%
^{2N}z_{0} \\
& =\left( \tfrac{1}{2\pi \hbar }\right) ^{N}U_{\phi }\left( \int (\mathcal{F}%
_{\sigma }A)(z_{0})\widehat{T}_{\text{Sch}}(z_{0})\psi )(z)\mathrm{d}%
^{2N}z_{0}\right) \\
& =U_{\phi }(\widehat{A}_{\text{Sch}}\psi )(z)
\end{align*}%
(the passage from the second equality to the third is legitimated by the
fact that $U_{\phi }$ is both linear and continuous).
\end{proof}

\section{Metaplectic Covariance\label{secmetacov}}

Since $Sp(N)$ is the symmetry group for the usual CCR (canonical commutation
relations)%
\begin{equation*}
\lbrack \widehat{X}_{j},\widehat{P}_{k}]=i\hbar \delta _{jk}
\end{equation*}%
for $\widehat{X}_{j}=x_{j}$, $\widehat{P}_{k}=-i\hbar \partial /\partial
x_{k}$), the uniqueness of \ these relations implies that for each $S$ in $%
Sp(N)$ there must be some associated unitary operator linking them to
quantization. These operators are the metaplectic operators.

\subsection{Metaplectic operators}

Let us recall how the metaplectic operators are defined (for details and
proofs see for instance\cite{Folland,AIF,ICP,Littlejohn}). Assume that $S$
is a free symplectic matrix, that is $S\in Sp(N)$ and 
\begin{equation*}
S=%
\begin{bmatrix}
A & B \\ 
C & D%
\end{bmatrix}%
\text{ \ \ with \ }\det B\neq 0\text{.}
\end{equation*}%
To every such $S$ one associates the operators $\pm \widehat{S}_{W,m}$
defined by the formula%
\begin{equation}
\widehat{S}_{W,m}\psi (x)=\left( \frac{1}{2\pi i\hbar }\right) ^{N/2}\frac{%
i^{m}}{\sqrt{|\det B|}}\int e^{\frac{i}{\hbar }W(x,x^{\prime })}\psi
(x^{\prime })\mathrm{d}^{N}x^{\prime }\text{;}  \label{qft}
\end{equation}%
here $W$ is Hamilton's characteristic function\ familiar from classical
mechanics (see for instance \cite{HGoldstein,ICP}): 
\begin{equation}
W(x,x^{\prime })=\frac{1}{2}DB^{-1}x^{2}-B^{-1}x\cdot x^{\prime }+\frac{1}{2}%
B^{-1}Ax^{\prime 2},  \label{wplq}
\end{equation}%
and $m$ is an integer (\textquotedblleft Maslov index\textquotedblright )
corresponding to a choice of $\arg \det B.$ The operators $\widehat{S}_{W,m}$
are a sort of generalized Fourier transform, and it is not difficult to
check that they are unitary. In addition the inverse of $\widehat{S}_{W,m}$
is given by%
\begin{equation*}
\widehat{S}_{W,m}^{-1}=\widehat{S}_{W^{\ast },m^{\ast }}\text{ \ , \ }%
W^{\ast }(x,x^{\prime })=-W(x^{\prime },x)\text{ \ , \ \ }m^{\ast }=N-m
\end{equation*}%
hence the $\widehat{S}_{W,m}$ generate a group: this group is the
metaplectic group $Mp(N)$ (see de Gosson \cite{ICP} for a complete
discussion of the properties of $Mp(N)$ and of the associated Maslov
indices). If we choose for $S$ the standard symplectic matrix%
\begin{equation*}
J=%
\begin{bmatrix}
0 & I \\ 
-I & 0%
\end{bmatrix}%
\end{equation*}%
the quadratic form (\ref{wplq}) reduces to $W(x,x^{\prime })=$ $-x\cdot
x^{\prime }$; choosing $\arg \det B=\arg \det I=0$ the corresponding
metaplectic operator is%
\begin{equation*}
\widehat{J}\psi (x)=\left( \frac{1}{2\pi i\hbar }\right) ^{N/2}\int e^{-%
\frac{i}{\hbar }x\cdot x^{\prime }}\psi (x^{\prime })\mathrm{d}^{N}x^{\prime
}=i^{-N/2}F\psi (x^{\prime })
\end{equation*}%
where $F$ thus is the usual unitary Fourier transform (\ref{ufo}).

The Wigner--Moyal transform enjoys the following metaplectic covariance
property: for every $\widehat{S}\in Mp(N)$ with projection $S\in Sp(N)$ we
have%
\begin{equation}
W(\widehat{S}\psi ,\widehat{S}\phi )=W(\psi ,\phi )\circ S^{-1}\text{.}
\label{metac1}
\end{equation}%
Since the Wigner wave-packet transform is defined in terms of $W(\psi ,\phi
) $ by formula (\ref{defwpt}) it follows that%
\begin{eqnarray*}
U_{\phi }(\widehat{S}\psi ) &=&\left( \frac{\pi \hbar }{2}\right) ^{N/2}W(%
\widehat{S}\psi ,\overline{\phi })(\tfrac{1}{2}z) \\
&=&\left( \frac{\pi \hbar }{2}\right) ^{N/2}W(\widehat{S}\psi ,\widehat{S}(%
\widehat{S}^{-1}\overline{\phi }))(\tfrac{1}{2}z) \\
&=&\left( \frac{\pi \hbar }{2}\right) ^{N/2}W(\psi ,\widehat{S}^{-1}%
\overline{\phi }))(\tfrac{1}{2}S^{-1}(z))
\end{eqnarray*}%
and hence 
\begin{equation}
U_{\phi }(\widehat{S}\psi )=(U_{\phi _{\widehat{S}}}\psi )\circ S^{-1}\text{
\ , \ }\phi _{\widehat{S}}=\overline{\widehat{S}^{-1}\overline{\phi }}\text{.%
}  \label{mcwpt}
\end{equation}

\subsection{Metaplectic group and Weyl calculus}

In \cite{MdGMW} we have shown, following an idea of Mehlig and Wilkinson 
\cite{WM}, that the metaplectic group is generated by operators of the type%
\begin{equation}
\widehat{S}\psi (x)=\left( \frac{1}{2\pi \hbar }\right) ^{N/2}\frac{i^{\nu
(S)}}{\sqrt{|\det (S-I)|}}\int e^{\frac{i}{2\hbar }M_{S}z_{0}^{2}}\widehat{T}%
_{\text{Sch}}(z_{0})\psi (x)\mathrm{d}^{2N}z_{0}  \label{MWO}
\end{equation}%
where $\det (S-I)\neq 0$, $M_{S}$ is the symplectic Cayley transform of $S$:%
\begin{equation*}
M_{S}=\frac{1}{2}J(S+I)(S-I)^{-1},
\end{equation*}%
and $\nu (S)$ is the Conley--Zehnder index (modulo $4$) of a path joining
the identity to $I$ in $Sp(N)$ (see for instance Muratore-Ginnaneschi \cite%
{MG} for a discussion of this index). For instance, if $\widehat{S}=\widehat{%
S}_{W,m}$ then%
\begin{equation}
\widehat{S}_{W,m}\psi (x)=\left( \frac{1}{2\pi \hbar }\right) ^{N/2}\frac{%
i^{m-\limfunc{Inert}W_{xx}}}{\sqrt{|\det (S-I)|}}\int e^{\frac{i}{2\hbar }%
M_{S}z_{0}^{2}}\widehat{T}_{\text{Sch}}(z_{0})\psi (x)\mathrm{d}^{2N}z_{0}
\label{MW1}
\end{equation}%
where $\limfunc{Inert}W_{xx}$ the number of negative eigenvalues of the
Hessian matrix of $W$. Formulae (\ref{MWO}) and (\ref{MW1}) are thus the
Weyl representations of the metaplectic operators $\widehat{S}$ and $%
\widehat{S}_{W,m}$). They allow us to define phase-space metaplectic
operators $\widehat{S}_{\text{ph}}$ in the following way: if $\det (S-I)\neq
0$ we set 
\begin{equation}
\widehat{S}_{\text{ph}}\Psi (z)=\left( \frac{1}{2\pi \hbar }\right) ^{N/2}%
\frac{i^{\nu (S)}}{\sqrt{|\det (S-I)|}}\int e^{\frac{i}{2\hbar }%
M_{S}z_{0}^{2}}\widehat{T}_{\text{ph}}(z_{0})\Psi (z)\mathrm{d}^{2N}z_{0}%
\text{;}  \label{meta}
\end{equation}%
the operators $\widehat{S}_{\text{ph}}$ are in one-to-one correspondence
with the metaplectic operators $\widehat{S}$ and thus generate a group which
we denote by $Mp_{\text{ph}}(N)$ (the \textquotedblleft phase space
metaplectic group\textquotedblright ). In following lemma we give
alternative descriptions of the operators (\ref{MWO}) in terms of the
operators $\widehat{T}_{\text{ph}}$:

\begin{lemma}
\label{lemma1}Let $\widehat{S}_{\text{ph}}\in Mp_{\text{ph}}(N)$ have
projection $S\in Mp(N)$ such that $\det (S-I)\neq 0$. We have 
\begin{equation}
\widehat{S}_{\text{ph}}=\left( \tfrac{1}{2\pi }\right) ^{N}i^{\nu (S)}\sqrt{%
|\det (S-I)|}\int e^{-\frac{i}{2}Sz\wedge z}\widehat{T}_{\text{ph}}((S-I)z)%
\mathrm{d}^{2N}z  \label{alf2}
\end{equation}%
and 
\begin{equation}
\widehat{S}_{\text{ph}}=\left( \tfrac{1}{2\pi }\right) ^{N}i^{\nu (S)}\sqrt{%
|\det (S-I)|}\int \widehat{T}_{\text{ph}}(Sz)\widehat{T}_{\text{ph}}(-z)%
\mathrm{d}^{2N}z\text{.}  \label{alf1}
\end{equation}
\end{lemma}

\begin{proof}
It is \textit{mutatis mutandis} the proof of Lemma 1 in \cite{MdGMW}): we
have 
\begin{equation*}
\tfrac{1}{2}J(S+I)(S-I)^{-1}=\tfrac{1}{2}J+J(S-I)^{-1}
\end{equation*}%
hence, in view of the antisymmetry of $J$,%
\begin{equation*}
M_{S}z\cdot z=J(S-I)^{-1}z\cdot z=(S-I)^{-1}z\wedge z
\end{equation*}%
Performing the change of variables $z\longmapsto (S-I)z$ we can rewrite the
integral in the right-hand side of (\ref{MW1}) as%
\begin{eqnarray*}
\int e^{\frac{i}{2}\left\langle M_{S}z,z\right\rangle }\widehat{T}(z)\mathrm{%
d}^{2N}z &=&\sqrt{|\det (S-I)|}\int e^{\frac{i}{2}z\wedge (S-I)z}\widehat{T}%
_{\text{ph}}((S-I)z)\mathrm{d}^{2N}z \\
&=&\sqrt{|\det (S-I)|}\int e^{-\frac{i}{2}Sz\wedge z}\widehat{T}_{\text{ph}%
}((S-I)z)\mathrm{d}^{2N}z
\end{eqnarray*}%
hence (\ref{alf2}). Taking into account formula (\ref{formuco3}) for the
product of two metaplectic operators $\widehat{T}_{\text{ph}}(z_{0})$ and $%
\widehat{T}_{\text{ph}}(z_{1})$ we get%
\begin{equation*}
\widehat{T}((S-I)z)=e^{\tfrac{i}{2}\sigma Sz\wedge z}\widehat{T}_{\text{ph}%
}(Sz)\widehat{T}_{\text{ph}}(-z)
\end{equation*}%
and formula (\ref{alf1}) follows.
\end{proof}

This result will allows us to show in a simple way that the well-known
\textquotedblleft metaplectic covariance\textquotedblright\ relation 
\begin{equation}
\widehat{A\circ S}_{\text{Sch}}=\widehat{S}^{-1}\widehat{A}_{\text{Sch}}%
\widehat{S}  \label{metacoweyl}
\end{equation}%
valid for any $\widehat{S}\in Mp(N)$ with projection $S\in Sp(N)$ extends to
the phase-space Weyl operators $\widehat{A}_{\text{ph}}$ provided one
replaces $Mp(N)$ with $Mp_{\text{ph}}(N)$.

\begin{theorem}
Let $S$ be a symplectic matrix and $\widehat{S}_{\text{ph}}$ any of the two
operators in $Mp_{\text{ph}}(N)$ associated with $S$. The following
phase-space metaplectic covariance formulae hold:%
\begin{equation}
\widehat{S}_{\text{ph}}\widehat{T}_{\text{ph}}(z_{0})\widehat{S}_{\text{ph}%
}^{-1}=\widehat{T}_{\text{ph}}(Sz)\text{ \ , \ }\widehat{A\circ S}_{\text{ph}%
}=\widehat{S}_{\text{ph}}^{-1}\widehat{A}_{\text{ph}}\widehat{S}_{\text{ph}}.
\label{metaco}
\end{equation}
\end{theorem}

\begin{proof}
To prove the first formula (\ref{metaco}) it is sufficient to assume that $%
\det (S-I)\neq 0$ and that $\widehat{S}_{\text{ph}}$is thus given by formula
(\ref{meta}): since such operators generate $Mp_{\text{ph}}(N)$. Let us thus
prove that 
\begin{equation}
\widehat{T}_{\text{ph}}(Sz_{0})\widehat{S}_{\text{ph}}=\widehat{S}_{\text{ph}%
}\widehat{T}_{\text{ph}}(z_{0})\text{ \ if \ }\det (S-I)\neq 0\text{.}
\label{44a}
\end{equation}%
Using either formula (\ref{alf1}) in Lemma \ref{lemma1} above and setting 
\begin{equation*}
C_{S}=\left( \tfrac{1}{2\pi }\right) ^{N}i^{\nu (S)}\sqrt{|\det (S-I)|}
\end{equation*}%
we have 
\begin{equation*}
\widehat{T}_{\text{ph}}(Sz_{0})\widehat{S}_{\text{ph}}=C_{S}\int \widehat{T}%
_{\text{ph}}(Sz_{0})\widehat{T}_{\text{ph}}(Sz)\widehat{T}_{\text{ph}}(-z)%
\mathrm{d}^{2N}z
\end{equation*}%
and%
\begin{equation*}
\widehat{S}_{\text{ph}}\widehat{T}_{\text{ph}}(z_{0})=C_{S}\int \widehat{T}_{%
\text{ph}}(Sz)\widehat{T}_{\text{ph}}(-z)\widehat{T}_{\text{ph}}(z_{0})%
\mathrm{d}^{2N}z.
\end{equation*}%
Setting%
\begin{equation*}
A(z_{0})=\int \widehat{T}_{\text{ph}}(Sz_{0})\widehat{T}_{\text{ph}}(Sz)%
\widehat{T}_{\text{ph}}(-z)\mathrm{d}^{2N}z
\end{equation*}%
and%
\begin{equation*}
B(z_{0})=\int \widehat{T}_{\text{ph}}(Sz)\widehat{T}_{\text{ph}}(-z)\widehat{%
T}_{\text{ph}}(z_{0})\mathrm{d}^{2N}z
\end{equation*}%
we have, by repeated use of (\ref{formuco3}), 
\begin{eqnarray*}
A(z_{0}) &=&\int e^{\frac{i}{2\hbar }\Phi _{1}(z,z_{0})}\widehat{T}_{\text{ph%
}}(Sz_{0}+(S-I)z)\mathrm{d}^{2N}z \\
B(z_{0}) &=&\int e^{\frac{i}{2\hbar }\Phi _{2}(z,z_{0})}\widehat{T}_{\text{ph%
}}(z_{0}+(S-I)z)\mathrm{d}^{2N}z
\end{eqnarray*}%
where the phases $\Phi _{1}$ and $\Phi _{2}$ are given by 
\begin{eqnarray*}
\Phi _{1}(z,z_{0}) &=&z_{0}\wedge z-S(z+z_{0})\wedge z \\
\Phi _{2}(z,z_{0}) &=&-Sz\wedge z+(S-I)z\wedge z_{0}\text{.}
\end{eqnarray*}%
Performing the change of variables $z^{\prime }=z+z_{0}$ in the integral
defining $A(z_{0})$ we get%
\begin{equation*}
A(z_{0})=\int e^{\frac{i}{2\hbar }\Phi _{1}(z^{\prime }-z_{0},z_{0})}%
\widehat{T}_{\text{ph}}(z_{0}+(S-I)z^{\prime })\mathrm{d}^{2N}z^{\prime }
\end{equation*}%
and%
\begin{eqnarray*}
\Phi _{1}(z^{\prime }-z_{0},z_{0}) &=&z_{0}\wedge (z^{\prime
}-z_{0})-Sz^{\prime }\wedge (z^{\prime }-z_{0}) \\
&=&(S-I)z^{\prime }\wedge z_{0}-Sz^{\prime }\wedge z^{\prime } \\
&=&\Phi _{2}(z^{\prime },z_{0})
\end{eqnarray*}%
hence (\ref{44a}). The second formula (\ref{metaco}) easily follows from the
first: noting that the symplectic Fourier transform (\ref{asigma}) satisfies 
\begin{eqnarray*}
\mathcal{F}_{\sigma }[A\circ S](z) &=&\left( \tfrac{1}{2\pi \hbar }\right)
^{N}\int e^{-\frac{i}{\hbar }z_{0}\wedge z^{\prime }}A(Sz^{\prime })\mathrm{d%
}^{2N}z^{\prime } \\
&=&\left( \tfrac{1}{2\pi \hbar }\right) ^{N}\int e^{-\frac{i}{\hbar }%
Sz_{0}\wedge z^{\prime }}A(z^{\prime })\mathrm{d}^{2N}z^{\prime } \\
&=&\mathcal{F}_{\sigma }A(Sz)
\end{eqnarray*}%
we have%
\begin{eqnarray*}
\widehat{A\circ S}_{\text{ph}} &=&\left( \tfrac{1}{2\pi \hbar }\right)
^{N}\int \mathcal{F}_{\sigma }A(Sz)\widehat{T}_{\text{ph}}(z)\mathrm{d}^{2N}z
\\
&=&\left( \tfrac{1}{2\pi \hbar }\right) ^{N}\int \mathcal{F}_{\sigma }A(z)%
\widehat{T}_{\text{ph}}(S^{-1}z)\mathrm{d}^{2N}z \\
&=&\left( \tfrac{1}{2\pi \hbar }\right) ^{N}\int \mathcal{F}_{\sigma }A(z)%
\widehat{S}_{\text{ph}}^{-1}\widehat{T}_{\text{ph}}(z)\widehat{S}_{\text{ph}}%
\mathrm{d}^{2N}z
\end{eqnarray*}%
which concludes the proof.
\end{proof}

It can be shown, adapting the proof of a classical result of Shale \cite%
{Shale} (see Wong \cite{Wong}, Chapter 30, for a proof) that the metaplectic
covariance formula 
\begin{equation*}
\text{\ }\widehat{A\circ S}_{\text{ph}}=\widehat{S}_{\text{ph}}^{-1}\widehat{%
A}_{\text{ph}}\widehat{S}_{\text{ph}}
\end{equation*}%
actually characterizes the phase-space Weyl operators $\widehat{A}_{\text{ph}%
}$. That is, any operator satisfying this relation for all operators $%
\widehat{S}_{\text{ph}}\in Mp_{\text{ph}}(N)$ is necessarily of the type%
\begin{equation*}
\widehat{A}_{\text{ph}}=\left( \tfrac{1}{2\pi \hbar }\right) ^{N}\int (%
\mathcal{F}_{\sigma }A)(z)\widehat{T}_{\text{ph}}(z)\mathrm{d}^{2N}z\text{.}
\end{equation*}

For example, if $H$ is the Hamiltonian function of the one-dimensional
harmonic oscillator put in normal form%
\begin{equation}
H=\frac{\omega }{2}(p^{2}+x^{2})  \label{oha1}
\end{equation}%
we get%
\begin{equation}
\widehat{H}_{\text{ph}}=-\frac{\hbar ^{2}\omega }{2}\nabla _{z}^{2}-i\frac{%
\hbar \omega }{2}z\wedge \nabla _{z}+\frac{\omega }{8}|z|^{2}  \label{oha2}
\end{equation}%
where $\nabla _{z}$ is the gradient operator in $(x,p)$.

\section{Schr\"{o}dinger Equation in Phase Space\label{sesch}}

We now have all the machinery needed to justify and stud the Schr\"{o}dinger
equation in phase space.

\subsection{The relationship between $\protect\psi $ and $\Psi $}

The following consequence of theorem \ref{thinter} links standard
\textquotedblleft configuration space\textquotedblright\ quantum mechanics
to phase-space quantum mechanics via the Wigner wave-packet transform and
the extended Heisenberg group studied in the previous sections. For clarity
we denote by $\widehat{A}_{\text{Sch}}$ the usual Weyl operator associated
by (\ref{weyl1}) to an observable $A$.

\begin{corollary}
\label{thufi}Let $U_{\phi }$, $\phi \in \mathcal{S}(\mathbb{R}_{x}^{N})$, be
an arbitrary Wigner wave-packet transform. (i) If $\psi =\psi (x,t)$ is a
solution of the usual Schr\"{o}dinger's equation%
\begin{equation*}
i\hbar \frac{\partial \psi }{\partial t}=\widehat{H}_{\text{Sch}}\psi
\end{equation*}%
then $\Psi =(U_{\phi }\psi )(z,t)$ is a solution of the phase-space Schr\"{o}%
dinger equation%
\begin{equation}
i\hbar \frac{\partial \Psi }{\partial t}=\widehat{H}_{\text{ph}}\Psi .
\label{phsch}
\end{equation}%
(ii) Assume that $\Psi $ is a solution of this equation and that $\Psi
_{0}=\Psi (\cdot ,0)$ belongs to the range $\mathcal{H}_{\phi }$ of $U_{\phi
}$. Then $\Psi (\cdot ,t)\in \mathcal{H}_{\phi }$ for every $t$ for which $%
\Psi $ is defined.
\end{corollary}

\begin{proof}
Since time-derivatives obviously commute with $U_{\phi }$ we have, using (%
\ref{inter})%
\begin{equation*}
i\hbar \frac{\partial \Psi }{\partial t}=U_{\phi }(\widehat{H}_{\text{Sch}%
}\psi )=\widehat{H}_{\text{ph}}(U_{\phi }\psi )=\widehat{H}_{\text{ph}}\Psi
\end{equation*}%
hence \textit{(i)}. Statement \textit{(ii)} follows.
\end{proof}

The result above leads to the following interesting questions: since the
solutions of the phase-space Schr\"{o}dinger equation (\ref{phsch}) exist
independently of the choice of any isometry $U_{\phi }$, what is the
difference in the physical interpretations of the corresponding
configuration-space wavefunctions $\psi =U_{\phi }^{\ast }\Psi $ and $\psi
^{\prime }=U_{\phi ^{\prime }}^{\ast }\Psi $? The answer is that there is no
difference at all provided that $\phi $ and $\phi ^{\prime }$ are not
orthogonal:

\begin{theorem}
Let $\Psi $ be a solution of the phase space Schr\"{o}dinger equation (\ref%
{phsch}) with initial condition $\Psi _{0}$ and define functions $\psi _{1}$
and $\psi _{2}$ in $L(\mathbb{R}_{x}^{N})$ by 
\begin{equation*}
\Psi =U_{\phi _{1}}\psi _{1}=U_{\phi _{2}}\psi _{2}\text{.}
\end{equation*}%
We assume that $\Psi _{0}\in \mathcal{H}_{\phi _{1}}\cap \mathcal{H}_{\phi
_{2}}$;

(i) We have $\Psi (\cdot ,t)\in \mathcal{H}_{\phi _{1}}\cap \mathcal{H}%
_{\phi _{2}}$ for all $t$

(ii) If $(\phi _{1},\phi _{2})=0$ then $\psi _{1}$ and $\psi _{2}$ are
orthogonal quantum states: $(\psi _{1},\psi _{2})=0$.
\end{theorem}

\begin{proof}
Property \textit{(i)} follows from Theorem \ref{thufi}\textit{(iii)}. Let us
prove \textit{(ii)}. In view of formula (\ref{important}) we have 
\begin{equation*}
((U_{\phi _{1}}\psi _{1},U_{\phi _{2}}\psi _{2}))=(\psi _{1},\psi _{2})%
\overline{(\overline{\phi }_{1},\overline{\phi }_{2})}
\end{equation*}%
that is $||\Psi |||^{2}=\lambda (\psi _{1},\psi _{2})$ with\ $\lambda =(\phi
_{1},\phi _{2})$. The assertion follows.
\end{proof}

\subsection{Spectral properties}

The operators $\widehat{A}_{\text{ph}}$ defined by (\ref{weyl2}) enjoy the
same property which makes the main appeal of ordinary Weyl operators, namely
that they are self-adjoint if and only if their symbols is real.

\begin{theorem}
Let $\widehat{A}_{\text{ph}}$ and $\widehat{A}_{\text{Sch}}$ be the
operators associated to a symbol $A$. We assume that the symplectic Fourier
transform $\mathcal{F}_{\sigma }A$ is defined. (i) The operator $\widehat{A}%
_{\text{ph}}$ is self-adjoint in $L^{2}(\mathbb{R}_{z}^{2N})$ if and only if 
$A=\overline{A}$; (ii) Every eigenvalue of $\widehat{A}_{\text{Sch}}$ is
also an eigenvalue of $\widehat{A}_{\text{ph}}$.
\end{theorem}

\begin{proof}
\textit{(i)} By definition of $\widehat{A}_{\text{ph}}$ and $\widehat{T}_{%
\text{ph}}$ we have%
\begin{eqnarray*}
\widehat{A}_{\text{ph}}\Psi (z) &=&\left( \tfrac{1}{2\pi \hbar }\right)
^{N}\int \mathcal{F}_{\sigma }A(z_{0})e^{-\frac{i}{2\hbar }z\wedge
z_{0}}\Psi (z-z_{0})\mathrm{d}^{2N}z_{0} \\
&=&\left( \tfrac{1}{2\pi \hbar }\right) ^{N}\int \mathcal{F}_{\sigma
}A(z-z^{\prime })e^{\frac{i}{2\hbar }z\wedge z^{\prime }}\Psi (z^{\prime })%
\mathrm{d}^{2N}z^{\prime }
\end{eqnarray*}%
hence the kernel of the operator $\widehat{A}_{\text{ph}}$ is%
\begin{equation*}
K(z,z^{\prime })=\left( \tfrac{1}{2\pi \hbar }\right) ^{N}e^{\frac{i}{2\hbar 
}z\wedge z^{\prime }}\mathcal{F}_{\sigma }A(z-z^{\prime })\text{.}
\end{equation*}%
In view of the standard theory of integral operators $\widehat{A}_{\text{ph}%
} $ is self-adjoint if and only if $K(z,z^{\prime })=\overline{K(z^{\prime
},z)}$; in view of the antisymmetry of the symplectic product we have%
\begin{equation*}
\overline{K(z^{\prime },z)}=\left( \tfrac{1}{2\pi \hbar }\right) ^{N}e^{%
\frac{i}{2\hbar }z\wedge z^{\prime }}\overline{\mathcal{F}_{\sigma
}A(z^{\prime }-z)}
\end{equation*}%
hence our claim since by definition (\ref{asigma}) of the symplectic Fourier
transform 
\begin{equation*}
\overline{\mathcal{F}_{\sigma }A(z^{\prime }-z)}=\left( \tfrac{1}{2\pi \hbar 
}\right) ^{N}\dint e^{-\frac{i}{\hbar }(z-z^{\prime })\wedge z^{\prime
\prime }}\overline{A(z^{\prime \prime })}\mathrm{d}^{2N}z^{\prime \prime }=%
\mathcal{F}_{\sigma }\overline{A}(z-z^{\prime }).
\end{equation*}%
\textit{(ii) }Assume that $\widehat{A}_{\text{Sch}}\psi =\lambda \psi $;
choosing $\phi \in \mathcal{S}(\mathbb{R}_{x}^{N})$ we have, using the
intertwining formula (\ref{inter}), 
\begin{equation*}
U_{\phi }(\widehat{A}_{\text{Sch}}\psi )=\widehat{A}_{\text{ph}}(U_{\phi
}\psi )=\lambda U_{\phi }\psi
\end{equation*}%
hence $\lambda $ is an eigenvalue of $\widehat{A}_{\text{ph}}$.
\end{proof}

Notice that there is no reason for an arbitrary eigenvalue of $\widehat{A}_{%
\text{ph}}$ to be an eigenvalue of $\widehat{A}_{\text{Sch}}$; this is only
the case if the corresponding eigenvector belongs to the range of a Wigner
wave-packet transform.

\subsection{The case of quadratic Hamiltonians}

There is an interesting application of the theory of the metaplectic group
outlined in Section \ref{secmetacov} to Schr\"{o}dinger's equation in phase
space. Assume that $H$ is a a quadratic Hamiltonian (for instance the
harmonic oscillator Hamiltonian); the flow determined by the associated
Hamilton equations is linear and consists of symplectic matrices $S_{t}$.
Letting time vary, thus obtain a curve $t\longmapsto S_{t}$ in the
symplectic group $Sp(N)$ passing through the identity $I$ at time $t=0$;
following general principles to that curve we can associate (in a unique
way) a curve $t\longmapsto \widehat{S}_{t}$ of metaplectic operators. Let
now $\psi _{0}=\psi _{0}(x)$ be some square integrable function and set $%
\psi (x,t)=\widehat{S}_{t}\psi _{0}(x)$. Then $\psi $ is just the solution
of the standard Schr\"{o}dinger's equation 
\begin{equation}
i\hbar \frac{\partial \psi }{\partial t}=\widehat{H}\psi \text{ \ \ , \ \ }%
\psi (t=0)=\psi _{0}  \label{schrq}
\end{equation}%
associated to the quadratic Hamiltonian function $H$. (Equivalently, $%
\widehat{S}_{t}$ is just the propagator for (\ref{schrq}).) This observation
allows us to solve explicitly the phase-space Schr\"{o}dinger equation for
any such $H$. Here is how. Since the wave-packet transform $U$ automatically
takes the solution $\psi $ of (\ref{schrq}) to a solution of the phase-space
Schr\"{o}dinger equation%
\begin{equation*}
i\hbar \frac{\partial \Psi }{\partial t}=\widehat{H}_{\text{ph}}\Psi
\end{equation*}%
we have%
\begin{equation*}
\Psi (z,t)=(\widehat{S}_{t})_{\text{ph}}\Psi (z,0)\text{.}
\end{equation*}%
Assume now that the symplectic matrix $S_{t}$ is free and $\det
(S_{t}-I)\neq 0$. Then, by (\ref{meta}),%
\begin{equation}
\Psi (z,t)=\left( \frac{1}{2\pi \hbar }\right) ^{N/2}\frac{i^{m(t)-\limfunc{%
Inert}W_{xx}(t)}}{\sqrt{|\det (S_{t}-I)|}}\int e^{\frac{i}{2\hbar }%
z_{0}^{T}M_{S}(t)z_{0}}\widehat{T}_{\text{ph}}(z_{0})\Psi (z,0)\mathrm{d}%
^{2N}z_{0}  \label{solution}
\end{equation}%
where $m(t)$, $W_{xx}(t)$, and $M_{S}(t)$ correspond to $S_{t}$. If $t$ is
such that $S_{t}$ is not free, or $\det (S_{t}-I)=0$, then it suffices to
write the propagator $\widehat{S}_{t}$ as the product of two operators (\ref%
{MW1}); note however that such values of $t$ are exceptional, and that the
solution (\ref{solution}) can be extended by taking the limit near such $t$
provided that takes some care in calculating the Maslov indices.

Let us illustrate this when $H$\ is the harmonic oscillator Hamiltonian
function (\ref{oha1}). The one-parameter group $(S_{t})$ is in this case
given by 
\begin{equation*}
S_{t}=%
\begin{bmatrix}
\cos \omega t & \sin \omega t \\ 
-\sin \omega t & \cos \omega t%
\end{bmatrix}%
\end{equation*}%
and the Hamilton principal function by%
\begin{equation*}
W(x,x^{\prime };t)=\frac{1}{2\sin \omega t}((x^{2}+x^{\prime 2})\cos \omega
t-2xx^{\prime }).
\end{equation*}%
A straightforward calculation yields%
\begin{equation*}
M_{S}(t)=%
\begin{bmatrix}
\frac{\sin \omega t}{-2\cos \omega t+2} & 0 \\ 
0 & \frac{\sin \omega t}{-2\cos \omega t+2}%
\end{bmatrix}%
=\frac{1}{2}%
\begin{bmatrix}
\cot (\frac{\omega t}{2}) & 0 \\ 
0 & \cot (\frac{\omega t}{2})%
\end{bmatrix}%
\end{equation*}%
and 
\begin{equation*}
\det (S_{t}-I)=2(1-\cos \omega t)=4\sin ^{2}(\tfrac{\omega t}{2});
\end{equation*}%
moreover%
\begin{equation*}
W_{xx}(t)=-\tan (\tfrac{\omega t}{2})\text{.}
\end{equation*}%
Insertion in formula (\ref{solution}) yields the explicit solution%
\begin{equation*}
\Psi (z,t)=\frac{i^{\nu (t)}}{2\left\vert 2\pi \hbar \sin (\tfrac{\omega t}{2%
})\right\vert ^{1/2}}\int \exp \left[ \frac{i}{4\hbar }(x_{0}^{2}+p_{0}^{2})%
\cot (\frac{\omega t}{2})\right] \widehat{T}_{\text{ph}}(z_{0})\Psi (z,0)%
\mathrm{d}^{2}z_{0}
\end{equation*}%
with 
\begin{equation*}
\nu (t)=\left\{ 
\begin{array}{c}
0\text{ \ if \ }0<t<\frac{\pi }{\omega } \\ 
-2\text{ \ if \ }-\frac{\pi }{\omega }<t<0\text{.}%
\end{array}%
\right.
\end{equation*}

\section{Interpretation of the phase-space wavefunction $\Psi $\label%
{secasol}}

Let us shortly discuss the probabilistic interpretation of the solutions $%
\Psi $ of the phase-space Schr\"{o}dinger equation%
\begin{equation*}
i\hbar \frac{\partial \Psi }{\partial t}=\widehat{H}_{\text{ph}}\Psi ;
\end{equation*}%
we will in particular elucidate the role played by $\phi $.

\subsection{Marginal probabilities}

Let $\psi $ be in $L^{2}(\mathbb{R}_{x}^{N})$; if $\psi $ is normalized: $%
||\psi ||=1$ then so is $\Psi =U_{\phi }\psi $ in view of the Parseval
formula (\ref{parseval}): $|||\Psi |||=1$. It follows that $|\Psi |^{2}$ is
a probability density in phase space. It turns out that by an appropriate
choice of $\phi $ the marginal probabilities can be chosen arbitrarily close
to $|\psi |^{2}$ and $|F\psi |^{2}$:

\begin{theorem}
\label{thepro}Let $\psi \in L^{2}(\mathbb{R}_{x}^{N})$ and set $\Psi
=U_{\phi }\psi $. (i) We have 
\begin{align}
\int |\Psi (x,p)|^{2}\mathrm{d}^{N}p& =(|\phi |^{2}\ast |\psi |^{2})(x)
\label{wpconvo1} \\
\int |\Psi (x,p)|^{2}\mathrm{d}^{N}x& =(|F\phi |^{2}\ast |F\psi |^{2})(p).
\label{wpconvo2}
\end{align}%
(ii) Let $\left\langle A\right\rangle _{\psi }=(A_{\text{Sch}}\psi ,\psi )$
be the mathematical expectation of the symbol $A$ in the normalized quantum
state $\psi $. We have%
\begin{equation}
\left\langle A\right\rangle _{\psi }=((A_{\text{ph}}\Psi ,\Psi ))\text{ \ ,
\ }\Psi =U_{\phi }\psi \text{.}  \label{ave}
\end{equation}
\end{theorem}

\begin{proof}
We have, by definition of $\Psi $,%
\begin{equation*}
|\Psi (z)|^{2}=\left( \frac{1}{2\pi \hbar }\right) ^{N}\iint e^{-\frac{i}{%
\hbar }p\cdot (x^{\prime }-x^{\prime \prime })}\overline{\phi }(x-x^{\prime
})\phi (x-x^{\prime \prime })\psi (x^{\prime })\overline{\psi }(x^{\prime
\prime })\mathrm{d}^{N}x^{\prime }\mathrm{d}^{N}x^{\prime \prime }\text{.}
\end{equation*}%
Since we have, by the Fourier inversion formula,%
\begin{equation*}
\int e^{-\frac{i}{\hbar }p\cdot (x^{\prime }-x^{\prime \prime })}\mathrm{d}%
^{N}p=(2\pi \hbar )^{N}\delta (x^{\prime }-x^{\prime \prime })
\end{equation*}%
it follows that%
\begin{align*}
\int |\Psi (z)|^{2}\mathrm{d}^{N}p& =\iiint \delta (x^{\prime }-x^{\prime
\prime })|\phi (x-x^{\prime })|^{2}|\psi (x^{\prime })|^{2}\mathrm{d}%
^{N}x^{\prime }\mathrm{d}^{N}x^{\prime \prime } \\
& =\int \left[ \int \delta (x^{\prime }-x^{\prime \prime })\mathrm{d}%
^{N}x^{\prime \prime }\right] |\phi (x-x^{\prime })|^{2}|\psi (x^{\prime
})|^{2}\mathrm{d}^{N}x^{\prime } \\
& =\int |\phi (x-x^{\prime })|^{2}|\psi (x^{\prime })|^{2}\mathrm{d}%
^{N}x^{\prime }
\end{align*}%
hence formula (\ref{wpconvo1}). To prove (\ref{wpconvo2}) we note that in
view of the metaplectic covariance formula (\ref{mcwpt}) for the wavepacket
transform we have%
\begin{equation*}
U_{\widehat{J}\phi }(\widehat{J}\psi )(x,p)=U_{\phi }\psi (-p,x)
\end{equation*}%
where $\widehat{J}=i^{-N/2}F$ is the metaplectic Fourier transform. It
follows that 
\begin{equation*}
U_{F\phi }(F\psi )(x,p)=i^{-N}U_{\phi }\psi (-p,x)
\end{equation*}%
and hence changing $(-p,x)$ into $(x,p)$:%
\begin{equation*}
U_{\phi }\psi (x,p)=i^{N}U_{F\phi }(F\psi )(p,-x)\text{.}
\end{equation*}%
and hence, using (\ref{wpconvo1}),%
\begin{align*}
\int |\Psi (x,p)|^{2}\mathrm{d}^{N}x& =\int |U_{F\phi }(F\psi )(p,-x)|^{2}%
\mathrm{d}^{N}x \\
& =\int |U_{F\phi }(F\psi )(p,x)|^{2}\mathrm{d}^{N}x \\
& =(|F\phi |^{2}\ast |F\psi |^{2})(p)
\end{align*}%
which concludes the proof of (\ref{wpconvo2}). To prove (\ref{ave}) it
suffices to note that in view of the intertwining formula (\ref{inter}) and
the fact that $U_{\phi }^{\ast }=U_{\phi }^{-1}$ we have 
\begin{eqnarray*}
((A_{\text{ph}}\Psi ,\Psi )) &=&((\widehat{A}_{\text{ph}}U_{\phi }\psi
,U_{\phi }\psi )) \\
&=&(U_{\phi }^{\ast }\widehat{A}_{\text{ph}}U_{\phi }\psi ,\psi ) \\
&=&(\widehat{A}_{\text{Sch}}\psi ,\psi )\text{.}
\end{eqnarray*}
\end{proof}

The result above shows that the marginal probabilities of $|\Psi |^{2}$ are
just the traditional position and momentum probability densities $|\psi
|^{2} $ and $|F\psi |^{2}$ \textquotedblleft smoothed out\textquotedblright\
by convoluting them with $|\phi |^{2}$ and $|F\phi |^{2}$ respectively.

\subsection{The limit $\hbar \rightarrow 0$}

Assume now that we choose for $\phi $ the Gaussian (\ref{fizero}): 
\begin{equation*}
\phi (x)=\phi _{\hslash }(x)=\left( \tfrac{1}{\pi \hbar }\right) ^{N/4}\exp
\left( -\frac{1}{2\hbar }|x|^{2}\right) \text{.}
\end{equation*}%
The Fourier transform of $\phi $ is identical to $\phi $ 
\begin{equation*}
F\phi _{\hbar }(p)=\left( \tfrac{1}{\pi \hbar }\right) ^{N/4}\exp \left( -%
\frac{1}{2\hbar }|p|^{2}\right) =\phi _{\hbar }(p)
\end{equation*}%
hence, setting $\Psi _{\hbar }=U_{\phi _{\hbar }}\psi $, and observing that $%
|\phi _{\hslash }|^{2}\rightarrow \delta $ when $\hbar \rightarrow 0$: 
\begin{align*}
\int |\Psi _{\hbar }(x,p)|^{2}\mathrm{d}^{N}p& =(|\psi |^{2}\ast |\phi
_{\hbar }|^{2})(x)\overset{\hbar \rightarrow 0}{\longrightarrow }|\psi
(x)|^{2} \\
\int |\Psi _{\hbar }(x,p)|^{2}\mathrm{d}^{N}x& =(|F\psi |^{2}\ast |\phi
_{\hbar }|^{2})(p)\overset{\hbar \rightarrow 0}{\longrightarrow }|F\psi
(p)|^{2}\text{.}
\end{align*}%
Thus, in the limit $\hbar \rightarrow 0$ the square of the modulus of the
phase-space wavefunction becomes a true joint probability density for the
probability densities $|\psi |^{2}$ and $|F\psi |^{2}$.

\section{Discussion and Remarks}

We have exposed some theoretical background for a mathematical justification
of the phase-space Schr\"{o}dinger equation%
\begin{equation*}
i\hbar \frac{\partial }{\partial t}\Psi (x,p,t)=H\left( \tfrac{1}{2}x+i\hbar 
\tfrac{\partial }{\partial p},\tfrac{1}{2}p-i\hbar \tfrac{\partial }{%
\partial x}\right) \Psi (x,p,t)\text{.}
\end{equation*}%
The aesthetic appeal of this equation is obvious --at least if one likes the
Hamiltonian formulation of mechanics. But is this equation \textit{useful}?
While the notion of \textquotedblleft usefulness\textquotedblright\ in
Science more than often has a relative and subjective character, one of the
main practical appeal of the phase-space Schr\"{o}dinger equation is that it
governs the quantum evolution of both pure and mixed states, while the
solutions of the usual Schr\"{o}dinger equation are, by definition, only
pure states. Another of the advantages of the phase-space approach is, as
pointed out in \cite{MMP}, the availability of factorization methods for the
Hamiltonian, for instance SUSY. From a practical point of view it could be
held against Schr\"{o}dinger equations in $2N$-dimensional phase space that
they are uninteresting because they involve solving a partial differential
equation in $2N+1$ variables instead of $N+1$ as for the ordinary Schr\"{o}%
dinger equation. But this is perhaps a somewhat stingy reservation
especially in times where modern computing techniques allow an efficient
processing of large strings of independent variables.

It would perhaps be interesting to make explicit the relationship between
the theory of Schr\"{o}dinger equation in phase space we have sketched and
other approaches to quantum mechanics in phase space, for instance the
\textquotedblleft deformation quantization\textquotedblright\ of Bayen 
\textit{et al.} \cite{BFFLS}, and whose master equation is the
\textquotedblleft quantum Liouville equation\textquotedblright .

We would like to end this section --and paper!-- by discussing a little bit
the possible physical interpretation of the phase space Schr\"{o}dinger
equation. Recall that we showed in Theorem \ref{theogauss} that a
phase-space Gaussian%
\begin{equation*}
\Psi _{G}(z)=\exp \left( -\frac{1}{2\hbar }Gz^{2}\right) \text{ \ , \ }%
G=G^{T}>0
\end{equation*}%
is in the range of any of the Wigner wave-packet transforms $U_{\phi }$ if
and only if $G\in Sp(N)$, and that in this case 
\begin{equation*}
W\psi (z)=|\Psi _{G}(z)|^{2}=\exp \left( -\frac{1}{\hbar }Gz^{2}\right)
\end{equation*}%
for some (pure) Gaussian state $\psi $. Let us more generally consider
Gaussians%
\begin{equation*}
\Psi _{M}(z)=\exp \left( -\frac{1}{\hbar }Mz^{2}\right)
\end{equation*}%
where $M$ is a positive-definite symmetric real matrix. One proves that $%
\Psi _{M}$ is the Wigner transform $W(\hat{\rho})$ of a (usually mixed)
quantum state if and only if $M^{-1}+iJ$ is positive-definite and Hermitian:%
\begin{equation}
(M^{-1}+iJ)^{\ast }=M^{-1}+iJ\geq 0\text{.}  \label{cond1}
\end{equation}%
The probabilistic meaning of this condition is the following: defining as
usual the covariance matrix of the state $\hat{\rho}$ by%
\begin{equation*}
\Sigma =\frac{\hbar }{2}M^{-1}
\end{equation*}%
condition (\ref{cond1}) can be rewritten as 
\begin{equation}
(\Sigma +i\frac{\hbar }{2}J)^{\ast }=\Sigma +i\frac{\hbar }{2}J\geq 0
\label{cond2}
\end{equation}%
which is equivalent to the uncertainty principle of quantum mechanics (see 
\cite{RS1,RS2} ; we have also discussed this in \cite{OPT}) . For instance,
when $N=1$ the matrix%
\begin{equation*}
\Sigma =%
\begin{bmatrix}
\Delta x^{2} & \Delta (x,p) \\ 
\Delta (x,p) & \Delta p^{2}%
\end{bmatrix}%
\end{equation*}%
satisfies (\ref{cond2}) if and only if%
\begin{equation*}
\Delta x^{2}\Delta p^{2}\geq \frac{1}{4}\hbar ^{2}+\Delta (x,p)
\end{equation*}%
which is the form of the Heisenberg inequality that should be used as soon
as correlations are present, and not the usual text-book inequality $\Delta
x\Delta p\geq \frac{1}{2}\hbar $.

It turns out that conditions (\ref{cond1})--(\ref{cond2}) have a simple
topological interpretation: we have shown in previous work of ours \cite%
{OPT,CQ} that they are equivalent to the third condition:

The phase-space ball $B(\sqrt{\hbar }):|z|\leq \hbar $ can be embedded into
the \textquotedblleft Wigner ellipsoid\textquotedblright\ $W_{M}:Mz^{2}\leq
\hbar $ using symplectic transformations (linear or not). Equivalently: the
symplectic capacity (or \textquotedblleft Gromov width\textquotedblright )
of $W_{M}$ is at least $\pi \hbar =\frac{1}{2}h$, one half of the quantum of
action: $c(W_{M})\geq \frac{1}{2}h$.

We have discussed in some detail in \cite{OPT,CQ} how this result allows a
\textquotedblleft coarse graining\textquotedblright\ of phase space by
symplectic quantum cells, which we dubbed \textquotedblleft quantum
blobs\textquotedblright . It appears that it is precisely this
coarse-graining that prevents Gaussians $\Psi _{M}$ with Wigner ellipsoids
smaller than a \textquotedblleft quantum blob\textquotedblright\ to
represent a quantum state. Is this to say that if the Wigner ellipsoid of $%
\Psi _{M}$ has exactly symplectic capacity $\frac{1}{2}h$ then $\Psi _{M}$
is a pure state? No, because such states are characterized by the fact that
the associated Wigner ellipsoid is exactly the image of the ball $B(\sqrt{%
\hbar })$ by a symplectic transformation since they are described by the
inequality $Gz^{2}=(Sz)^{2}\leq \hbar $ in view of Theorem \ref{theogauss},
and there are infinitely many ellipsoids with symplectic capacity $\frac{1}{2%
}h$ which are not the image of $B(\sqrt{\hbar })$ by a symplectic
transformation. However, we have shown in \cite{OPT,CQ} that if the
ellipsoid $W_{M}:Mz^{2}\leq \hbar $ has symplectic capacity $\frac{1}{2}h$
then one can associate to $W_{M}$ a unique pure Gaussian state. The argument
goes as follows: if $c(W_{M})=\frac{1}{2}h$ then if $S$ and $S^{\prime }$ in 
$Sp(N)$ are such that%
\begin{equation*}
S(B(\sqrt{\hbar }))\subset W_{M}\text{ \ , \ }S^{\prime }(B(\sqrt{\hbar }%
))\subset W_{M}\text{ }
\end{equation*}%
then there exists $R\in U(N)=Sp(N)\cap O(2N)$ such that $S=RS^{\prime }$
(the proof of this property is not entirely trivial) and hence $%
S^{T}S=(S^{\prime })^{T}S^{\prime }$. It follows that the ellipsoid $%
W_{G}:Gz^{2}\leq \hbar $, $G=S^{T}S$, is uniquely determined by $W_{M}$ and
that the pure Gaussian state corresponding to%
\begin{equation*}
\Psi _{G}(z)=\exp \left( -\frac{1}{2\hbar }Gz^{2}\right)
\end{equation*}%
is canonically associated to the mixed state $\Psi _{M}$, which does not in
general belong to the range of any Wigner wave-packet transform $U_{\phi }$.
It would be interesting to generalize this result to arbitrary functions $%
\Psi \in L^{2}(\mathbb{R}_{z}^{2N})$ by defining, in analogy with the
Gaussian case, a \textquotedblleft Wigner set\textquotedblright\ $W_{\Psi }$
associated with $\Psi $. One could then perhaps prove that $\Psi $
represents an arbitrary (mixed) quantum state provided that $W_{\Psi }$ has
a symplectic capacity at least $\frac{1}{2}h$. But enough is enough! We hope
to come back to these possibilities in the future. \bigskip

\noindent \textbf{Acknowledgement}. I wish to thank both referees for having
pointed out misprints and typos. This work has been partially supported by a
grant of the Max-Planck-Gesellschaft (Albert-Einstein-Institute, Golm). It
is a pleasure for me to thank Professor Dr. Hermann Nicolai for his warm
hospitality during part of the winter 2005 in Brandenburg. It has also been
supported by the FAPESP agency (Brazil) during the author's stay at the
University of S\~{a}o Paulo; I would like to thank Professor Paolo Piccione
for his generous invitation and for having provided a more than congenial
environment.

\end{document}